\documentclass[aps,prd,onecolumn,preprintnumbers,showpacs,nofootinbib,amssymb]{
revtex4}
\usepackage{graphicx}
\usepackage{amsmath}
\usepackage{amssymb}
\usepackage{amsfonts}
\usepackage{bm}
\usepackage{color}

\def\be{\begin{equation}}
\def\ee{\end{equation}}
\def\bea{\begin{eqnarray}}
\def\eea{\end{eqnarray}}

\begin{document}

\title{Statistical mechanics of self-gravitating systems in general
relativity:\\
II. The classical Boltzmann gas}
\author{Pierre-Henri Chavanis}
\affiliation{Laboratoire de Physique Th\'eorique, Universit\'e de Toulouse,
CNRS, UPS, France}

\begin{abstract} 

We study the statistical mechanics of classical self-gravitating systems
confined within a box of radius $R$  in general relativity. It has been
found that the caloric curve
$T_{\infty}(E)$ has the form of a double spiral whose
shape depends
on the compactness parameter $\nu=GNm/Rc^2$. The
double spiral shrinks as $\nu$ increases and finally 
disappears when
$\nu_{\rm max}=0.1764$. Therefore, general relativistic effects render the
system more unstable. On the other hand, the cold spiral
and the hot spiral move away from each other as $\nu$ decreases. Using a
normalization $\Lambda=-ER/GN^2m^2$ and $\eta=GNm^2/R k_B
T_{\infty}$
appropriate to the nonrelativistic limit, and considering 
$\nu\rightarrow 0$, the hot spiral goes to infinity and the caloric curve tends
towards a limit curve (determined by the Emden equation) exhibiting
a single cold spiral, as found in former
works. Using
another normalization ${\cal M}=GM/Rc^2$ and ${\cal
B}={Rc^4}/{GNk_B T_{\infty}}$ appropriate to the  ultrarelativistic limit, and
considering  $\nu\rightarrow 0$, 
the cold spiral goes to infinity and the caloric curve tends towards a limit
curve (determined by the general relativistic Emden equation) exhibiting a
single hot spiral. This result is new. We
discuss the analogies and the differences
between this asymptotic caloric curve and the caloric curve of the
self-gravitating black-body
radiation. Finally, we compare box-confined isothermal models with heavily
truncated isothermal distributions in Newtonian gravity and general
relativity.

\end{abstract}


\maketitle

\section{Introduction}

In a preceding paper \cite{theory1} (Paper I), relying on previous works on
the
subject
\cite{tolman,cocke,hk0,kh,khk,kam,kh2,hk,ipser80,sorkin,sy,sy2,bvrelat,aarelat1,
aarelat2,gao,gaoE,roupas1,gsw,fg,roupas1E,schiffrin,psw,fhj}, we
have developed
a general formalism to study the statistical mechanics of self-gravitating
systems in general relativity. This formalism is
valid for an arbitrary form of entropy. The statistical equilibrium state is
obtained by maximizing the entropy $S$ at fixed mass-energy $Mc^2$ and particle
number $N$.
The extremization problem
yields the Tolman-Oppenheimer-Volkoff (TOV) equations \cite{tolman,ov}
expressing the condition
of hydrostatic equilibrium together with the Tolman-Klein relations
\cite{tolman,klein}. The precise form of entropy determines the distribution
function and the equation of state of the system. For illustration, in Paper I,
we
considered a system of self-gravitating fermions. In that case, the
statistical equilibrium state is obtained by maximizing the Fermi-Dirac
entropy at fixed mass-energy and particle number. This yields the
relativistic Fermi-Dirac
distribution function.
In the present paper, we consider the statistical mechanics of
classical self-gravitating particles.\footnote{If the
particles are fermions, the classical limit corresponds to the nondegenerate
limit $T\gg T_F$  of the self-gravitating Fermi gas, where $T_F$ is the Fermi
temperature. This is what we will assume
here in order to make the connection with Paper I. However, our results are more
general since they can also describe a gas of bosons in the classical limit
$T\gg T_c$, where $T_c$ is the condensation temperature.} In that case, the
statistical equilibrium state is obtained by maximizing the Boltzmann
entropy at fixed mass-energy and particle number. This yields the
relativistic Maxwell-Boltzmann (or Maxwell-Juttner) distribution function. We
start by a brief history of the subject giving an exhaustive list of
references.\footnote{A detailed historic of the statistical
mechanics of
self-gravitating systems (classical and quantum) in Newtonian gravity and
general relativity is given in Refs. \cite{acf,acb,theory1}.}

The statistical mechanics of nonrelativistic classical  self-gravitating systems
was developed in connection with the dynamical evolution of globular clusters
(see Appendix D of \cite{clm1} for a short review).
Because of close two-body
gravitational encounters between stars, globular clusters have the tendency to
approach a
Maxwell-Boltzmann distribution \cite{chandrabook}. However, when coupled to
gravity, the isothermal
distribution  has an infinite mass so that it cannot be valid
everywhere in the cluster. In
practice, the relaxation towards the Maxwell-Boltzmann distribution is
hampered by the escape of high energy stars \cite{ambartsumian,spitzer} and the
Maxwell-Boltzmann distribution is established only in the core of the
system where stars have sufficiently negative energies. Since evaporation is a
slow process, globular clusters can be
found in quasiequilibrium states described by a truncated Maxwell-Boltzmann
distribution -- a Woolley distribution
\cite{woolley} or a King \cite{king}
distribution -- whose parameters slowly evolve in
time.\footnote{The Woolley \cite{woolley} distribution,
previously introduced by
Eddington \cite{eddington3}, corresponds to the isothermal distribution with
positive energies removed. The King \cite{king} distribution, previously
introduced by Michie \cite{michie} in a more general form, is a lowered
isothermal distribution such that the density of stars in phase space vanishes
continuously
above the escape energy.} The series of equilibria
$T(E)$ has the form of a spiral (see Fig. \ref{ebAintro}
of Appendix \ref{sec_scc}). Because of
close encounters and evaporation,
the system contracts and follows a series of equilibra $T(E)$ along
which the
energy $E$ decreases and the temperature $T$ increases (in the part of
the caloric curve where the specific
heat is negative) as the central density $\rho_0$
increases. The
thermodynamical
stability of the truncated Maxwell-Boltzmann distribution
(Woolley and King models) has been studied in \cite{lbw,katzking,clm1}. At the
turning
point of energy (minimum energy), when the
specific
heat vanishes, passing from negative to positive values, the series of
equilibria
becomes unstable and the system undergoes the Antonov instability
\cite{antonov}, also known as the gravothermal catastrophe \cite{lbw},
and collapses.
This instability can be followed by using dynamical 
models based either on moment equations derived from the
Fokker-Planck equation \cite{larson1,larson2}, Monte Carlo models
\cite{hmc}, heuristic fluid equations \cite{hnns,lbe}, or kinetic equations
such as the orbit-averaged-Fokker-Planck equation \cite{cohn}. During the
collapse the system takes a
core-halo structure in which the
cluster develops a dense and hot core and a diffuse envelope. The
dynamical evolution of the system is due to
the gradient of temperature (velocity dispersion) between the core and the halo
and the fact that the core has a negative specific heat. The core loses heat to
the profit of the halo, becomes hotter, and
contracts. If the temperature increases more rapidly in the
core than in the halo there is no possible equilibrium
and we get a thermal runaway: this is the gravothermal
catastrophe.
As a result, the core collapses
and reaches higher and higher densities and higher and higher temperatures while
the halo is not sensibly affected by the collapse of the
core and maintains its initial structure. The collapse of the core is
self-similar
and leads to a finite-time 
singularity: the central density and the temperature become infinite in
a finite time while the core radius shrinks to nothing \cite{lbe,cohn}. This is
called core
collapse. The mass contained in the core tends to zero at the
collapse time. The evolution may continue in a self-similar postcollapse regime
with the
formation of a binary star \cite{inagakilb}. The energy released by the binary
can
stop the collapse and induce a reexpansion of the system. Then, a series of
gravothermal oscillations is expected to follow \cite{oscillations,hr}. It has
to be noted that the gravothermal
catastrophe is a thermodynamical instability, not a dynamical instabilty.
Indeed, it
has
been shown that all the isotropic stellar systems with a distribution function
of
the form $f=f(\epsilon)$ with
$f'(\epsilon)<0$, like the truncated Maxwell-Boltzmann
distribution, are dynamically stable with respect to a collisionless
(Vlasov) evolution \cite{doremus71,doremus73,gillon76,sflp,ks,kandrup91}. In
particular, all the isothermal configurations on the series of equilibria are
dynamically stable, including those deep into the spiral that are
thermodynamically unstable. Therefore, dynamical and thermodynamical stability
do not coincide in Newtonian
gravity (see Paper I for a more detailed
discussion). As a result, the process of core collapse of
globular clusters is a very long (secular) process,
taking place on a collisional relaxation timescale of the order of the age of
the Universe since it is due to two-body gravitational encounters.

As the density and the temperature of the core increase during the gravothermal
catastrophe, the system 
may become relativistic. In that case, we have to consider the evolution of
general relativistic star clusters. The series of equilibria of heavily
truncated
Maxwell-Boltzmann distributions in general relativity (relativistic Woolley
model)
was first studied by Zel'dovich and Podurets
\cite{zp}.\footnote{Truncated isothermal distributions of
relativistic star clusters  have been studied independently by Fackerell
\cite{fackerellthesis}.}
They
discovered
the existence of a maximum temperature  $k_B (T_{\infty})_{\rm max}/mc^2=0.273$
in the
series of equilibria $T_{\infty}(\rho_0)$ as the central density
$\rho_0$, or the
central
redshift $z_0$, increases ($T_{\infty}$ is the temperature measured by an
observer at infinity). They argued that the system
should undergo a
gravitational collapse at that point that they called  an ``avalanche-type
catastrophic contraction of the system''. The mechanism of the collapse
proposed by Zeldovich and
Podurets \cite{zp} is the following. At the maximum temperature the orbits of
highly relativistic
particles become unstable and the corresponding particles start falling in
spirals towards the center. The
collapse of the orbits of some particles leads to an increase of the field
acting on the other particles, whose orbits collapse in turn etc. This
catastrophic collapse occurs rapidly, on a dynamical timescale.   A large
fraction of the system (the main mass)
 rapidly contracts to its gravitational radius and forms what is
now called a black hole (see
footnote 3 in paper I). However, only
the core of the system collapses. There remains a
cloud
surrounding the main mass.  The particles in the cloud, following the laws of
slow evolution, gradually fall into the collapsed mass.

 The
dynamical
\cite{fackerell,ipserthorne,ipser69,ipser69b,fackerell70,sudbury,sf76,fs76,
ipser80,mr,mreuro,mr90,bmrv,bmrv2} and
thermodynamical
\cite{hk0,kh,khk,kam,kh2,hk}
stability of heavily truncated Maxwell-Boltzmann (isothermal) distributions in
general
relativity have been studied by several authors.  Ipser \cite{ipser69b,ipser80}
showed that
dynamical and thermodynamical instabilities coincide in general relativity
and that they occur
at the turning point of the fractional binding energy $E/Nmc^2=(M-Nm)/Nm$,
corresponding to a central redshift $z_c=0.516$.\footnote{Ipser
\cite{ipser69b} (see also \cite{ipser69}) studied the dynamical
stability with respect to the Vlasov-Einstein equations of heavily truncated
isothermal distributions by using a variational principle based on the equation
of pulsations derived by Ipser and Thorne \cite{ipserthorne}. By using a
suitably chosen trial function he numerically obtained an approximate
expression of the square complex pulsation $\omega_{\rm app}^2$ which, by
construction, is always larger than the real one $\omega^2$. He showed that
$\omega_{\rm app}^2>0$ for $z<z_c\sim 0.5$ and $\omega_{\rm app}^2<0$ for
$z>z_c\sim
0.5$. The transition ($z_c\sim 0.5$) turns out to coincide with the turning
point of
fractional binding energy $E/Nmc^2$. This proves that the system becomes
unstable after
the turning point of binding energy and suggests (but does not prove) that the
system
is stable before the turning point of binding energy. The dynamical stability of
the system before the turning point of binding energy was proved later by Ipser
\cite{ipser80}. He first showed that thermodynamical stability
(in a very general
sense) implies dynamical stability. Then, using the Poincar\'e \cite{poincare}
criterion, he
showed (see also \cite{hk}) that the system is thermodynamically stable before
the turning point of binding energy and thermodynamically unstable after the
turning point of binding energy. This implies that the system is
dynamically stable before the turning point of binding energy. On the other
hand, since the  system is dynamically unstable after the turning point
of binding energy \cite{ipser69b}, Ipser \cite{ipser80}  concluded
that dynamical and thermodynamical stability  coincide in general
relativity. He conjectured that this equivalence between thermodynamical and
dynamical stability remains valid for all isotropic star clusters, not only
for those described by the heavily truncated Maxwell-Boltzmann distribution.
This is in sharp contrast with the Newtonian case where
it has been shown \cite{doremus71,doremus73,gillon76,sflp,ks,kandrup91} 
that  all
isotropic stellar systems are dynamically stable with respect to the
Vlasov-Poisson
equations, even those that are thermodynamically unstable. To solve this
apparent
paradox, one
expects that the growth rate $\lambda$ of the dynamical instability decreases
as relativity effects decrease and that it tends to zero in the
nonrelativistic limit $c\rightarrow +\infty$.} At
that point
$[(M-Nm)/Nm]_{\rm
c}=0.0357$, $(Rc^2/2GNm)_{\rm c}=4.42$, and $(k_BT_{\infty}/mc^2)_{\rm c}=0.23$.
This critical point 
is different from
the turning point of temperature reported by  Zel'dovich and Podurets \cite{zp}
corresponding to $z_0=1.08$, $(M-Nm)/Nm=0.0133$, $Rc^2/2GNm=3.92$, and
$k_B(T_{\infty})_{\rm max}/mc^2=0.27$. In particular, the gravitational
instability occurs sooner than predicted by  Zel'dovich and Podurets \cite{zp}. 
This led to the following scenario proposed by Fackerell {\it et
al.} \cite{fit} which refines the former scenario of  Zel'dovich and Podurets
\cite{zp}. They assumed that the evolution of relativistic spherical star
clusters in the nuclei of some galaxies (which may have resulted from the
gravothermal catastrophe of an initially Newtonian stellar
system)
proceeds quasistatically along a series of equilibria
$T_{\infty}(E)$.\footnote{Fackerell {\it et al.} \cite{fit}  worked in terms of
the curve $E(z_0)$ which presents damped oscillations (see Fig. 2 of
\cite{ipser69b}). The caloric curve $T_{\infty}(E)$
can be
obtained from Table I of 
Ipser \cite{ipser69b}. It is plotted in Fig. \ref{ipser} of Appendix
\ref{sec_scc} and has the form of a spiral.} The
evolution is driven
by stellar collisions and by the evaporation of stars. Both collision and
evaporation drive the cluster toward states of higher and higher central
density (or central redshift), higher and higher temperature (in the region
of negative specific heat) and lower and lower
binding energy. When the cluster reaches the turning point of fractional
binding energy (minimum fractional binding energy) it can no longer evolve
quasistatically and  it undergoes a dynamical
instability of general relativistic origin. Relativistic gravitational
collapse sets it: the stars spiral inward through the gravitational radius of
the cluster towards its center leaving behind a ``black hole'' in space with
some stars orbiting it.
Fackerell {\it et
al.} \cite{fit} speculated that violent events
in the nuclei of galaxies and in quasars might be associated with the onset of
such a collapse or with encounters between an already collapsed cluster (black
hole) and surrounding stars.

This scenario has been confirmed by Shapiro and Teukolsky
\cite{st1,st2,st3,st4,kochanek,rst1,strevue} who numerically
solved the relativistic Vlasov-Einstein equations
describing the dynamical evolution of a collisionless spherical gas of
particles in general relativity.
They specifically considered star clusters
made of compact stars such as white dwarfs, neutron stars or stellar mass black
holes. At the centers of the galaxies the collisions may
be sufficient to
induce a dynamical evolution.
Begining from a dense, but otherwise
Newtonian, star cluster, and following
secular core concentration on a two-body relaxation time scale (i.e. the
gravothermal catastrophe), the cluster may develop an extreme core-halo
configuration.  Therefore, if relativistic star
clusters form in nature they are
likely to be very centrally condensed. Because of collisions and evaporation,
the central density, the central redshift and the central velocity dispersion
increase. Shapiro and Teukolsky
\cite{st2} (see also \cite{rst1}) followed the series of equilibria
of truncated isothermal distributions and showed from direct numerical
simulations that above a critical redshift
$z_c\sim 0.5$, corresponding to the turning point of fractional binding energy,
the relativistic star cluster becomes dynamically unstable and undergoes a 
catastrophic collapse to a supermassive black hole on a dynamical time
scale.
Even in the case of extremely centrally condensed configurations
with extensive Newtonian halo, an appreciable fraction of the total cluster mass
collapses to the central black hole \cite{st4,kochanek}. This happens even
when the initial central core is just a small fraction of
the total mass.  This occurs because of the ``avalanche
effect'' predicted by Zel'dovich and Podurets \cite{zp} according to  which the
particles
in the cloud gradually fall into the collapsed
mass.\footnote{The gravitational
collapse of a supermassive star (fluid sphere) is a homologous radial infall of
all the fluid \cite{stgas1,stgas2}. By contrast, the gravitational collapse
of a collisionless relativistic star cluster is an inward spiralling of all
the stars \cite{st4}.}  Shapiro
and Teukolsky argued \cite{st3} that the
collapse of such clusters embedded in evolved galactic nuclei can lead to the
formation of supermassive
black holes of the ``right size'' ($10^6\le M/M_{\odot}\le 10^9$) to explain
quasars and active galactic
nuclei (AGNs).\footnote{Early works on relativistic star
clusters were stimulated by the scenario of Hoyle and Fowler
\cite{hf67} that quasars could be
supermassive star clusters (they had previously considered the possibility that
supermassive stars \cite{hf63a,hf63b} could be the
energy sources for quasars and
active galactic nuclei). In their theory, the observed high redshifts
of
quasars ($z\gtrsim 2$)
are explained by the fact that the clusters are very relativistic (gravitational
redshift) instead of being far away  (cosmological redshift). The fact that
all
relativistic star clusters studied by Ipser \cite{ipser69,ipser69b} and
Fackerell
\cite{fackerell70} were found to be unstable above
$z_c\sim 0.5$
rapidly threw doubts on their scenario \cite{zapolsky},
priviledging the scenario of Salpeter \cite{salpeter}
and Zel'dovich \cite{zeldovichBH} that supermassive black holes might be the
objects responsible for the energetic activity of quasars and galatic
nuclei. We now know that quasar redshifts
have a cosmological origin (they are far away). We note, however,
that there are examples of
relativistic star clusters that are stable at any redshift. A first example was
constructed by Bisnovatyi-Kogan and Zel'dovich \cite{bkz} (see also
Bisnovatyi-Kogan and Thorne \cite{bkt})  but this cluster is singular, having
infinite central density and infinite redshift, so it is
not very realistic (in addition the techniques
for testing stability yielded inconclusive results). Later, Rasio {\it et al.}
\cite{rst2} (see also Merafina and Ruffini \cite{mrz}) reported
a situation where the binding energy
has no turning point so that there is no dynamical instability. Specifically,
they followed the
gravothermal
catastrophe in the general relativistic regime and found that the binding energy
decreases monotonically with the central redshift so that the clusters remain
dynamically stable for all redshifts up to infinite central
redshift $z_0\rightarrow +\infty$. These
clusters could represent the relativistic final states of initially Newtonian
clusters undergoing the gravothermal catastrophe. {\it Remark--} According to
these results, it is not quite clear if the initial scenario proposed by 
Shapiro and Teukolsky \cite{st1,st2,st3,st4,kochanek,rst1,strevue} is correct.
Indeed, in their early works
\cite{st1,st2,st3,st4,kochanek,rst1,strevue}  they {\it assumed} that the
gravothermal
catastrophe transforms a Newtonian cluster into a relativistic one described by
a truncated isothermal distribution and then showed that this distribution
undergoes a dynamical instability of general relativistic origin and collapses
towards a black hole. However, in their later work (with Rasio) \cite{rst2} they
found that the relativistic generalisation of the distribution function
produced by the
gravothermal catastrophe \cite{cohn} differs from the
truncated isothermal
distribution and that it remains always dynamically stable up to infinite
central redshift. This seems to preclude the
formation of a black hole from the gravothermal catastrophe. This problem may
be solved by the refined scenario developed later by Balberg {\it et al.}
\cite{balberg}.}

In a later work, Balberg {\it et al.} \cite{balberg} (see also \cite{bash,pss})
applied similar ideas to self-interacting dark matter. Namely, they developed,
in the context of dark matter, the idea of ``avalanche-type contraction''
towards a supermassive black hole initially suggested by Zel'dovich and
Podurets \cite{zp}, improved by Fackerell {\it et al.} \cite{fit}, and
confirmed numerically  by Shapiro and Teukolsky
\cite{st2,st3,st4}.  They argued that dark matter halos experience
a gravothermal
catastrophe and that, when the central density and the temperature increase
above a critical value, the system
undergoes a dynamical instability of general relativistic
origin leading to the
formation of a supermassive black hole
on a
dynamical timescale. The dynamical evolution
of the system is due to the self-interaction of the
dark matter particles. In that case, a typical halo has sufficient time
to thermalize and acquire a gravothermal profile consisting of a flat core
surrounded by an extended halo. There is a first stage in which
the halo is in the long mean free path (LMFP) limit. It undergoes a
gravothermal catastrophe in which the core collapses self-similarly. This
is analogous (with slightly different exponents) to  the self-similar collapse 
of globular clusters. In this process, the
core mass decreases rapidly.
If this self-similar evolution were going to completion then, when the
core becomes relativistic, it would contain almost no mass. So, even if it
collapsed by a dynamical relativistic instability, it would not form a
supermassive black hole. But during the gravothermal catastrophe, the core of
self-interacting dark matter
halos passes in the  short mean free path (SMFP) limit. In the SMFP limit, the
core mass decreases more slowly (and almost saturates) so that a relatively
large mass can ultimately collapse into a supermassive black hole.  We note
that only the central region of a dark matter halo (not its outer
part) is affected by this process so the final outcome of this scenario is 
an isothermal halo harboring a central supermassive black hole. In recent
works, we have proposed to apply this scenario to isothermal models of dark
matter halos made of fermions \cite{clm1,clm2} or bosons \cite{bosons}.

In the previously mentioned studies, the star clusters are described by the
truncated Maxwell-Boltzmann distribution. In that case, the
link with standard thermodynamics is not straightforward because the truncated
Maxwell-Boltzmann distribution differs from the ordinary Maxwell-Boltzmann
distribution and describes an out-of-equilibrium situation where some stars
leave the system. Using the truncated
Maxwell-Boltzmann distribution  is compulsory if we want to
describe
realistic star clusters since the ordinary Maxwell-Boltzmann distribution
coupled to gravity has an infinite mass. However, it may be useful to study in
parallel
simplified models (somewhat academic) that correspond to standard statistical
mechanics based on the ordinary Maxwell-Boltzmann distribution. This can be
done by confining
artificially the
particles within a spherical box with reflecting boundary conditions in order to
have a finite mass.\footnote{The analogies  and the differences between
box-confined isothermal models and
heavily truncated isothermal distributions in Newtonian gravity and general
relativity are discussed in Appendix \ref{sec_sccnrb}. We note
that the presence of a box can substantially change the behavior of the caloric
curve and alter the physics of the problem. For nonrelativistic systems, the
caloric curve of box-confined isothermal systems is relatively similar to the
caloric
curve of the truncated isothermal (King or Woolley) model. However, for
general relativistic systems,  they are
very different from each other.} This ``box model'' was introduced by Antonov 
\cite{antonov} in the case of nonrelativistic stellar systems.
The statistical mechanics of nonrelativistic classical 
self-gravitating systems confined within a box has been studied by numerous
authors
\cite{antonov,lbw,ipser74,tvh,hkpart1,nakada,hs,hk3,katzpoincare1,ih,inagaki,
lecarkatz,ms,sflp,lp,kiessling,paddyapj,paddy,dvsc,semelin0,katzokamoto,semelin,
dvs1,dvs2,aaiso,crs,sc,grand,katzrevue,lifetime,ijmpb,sb}. These studies have
been
extended in
general
relativity by Katz and
Horwitz \cite{kh2} and more recently by Roupas \cite{roupas} and
Alberti and Chavanis \cite{acb}. The caloric curves (more precisely
the series of equilibria)  giving the normalized inverse temperature
$\eta=GNm^2/Rk_B T_{\infty}$  as
a function of the
normalized energy $\Lambda=-ER/GN^2m^2$, where $E=(M-Nm)c^2$ is the binding
energy, depend on a unique parameter $\nu=GNm/Rc^2$ called the compactness
parameter. It is found
that the caloric curves generically
present a double spiral. This
double spiral shrinks as $\nu$ increases and finally 
disappears at the  maximum compactness
$\nu_{\rm max}=0.1764$. Therefore, general relativistic
effects render the
system more unstable \cite{roupas,acb}.

The ``cold spiral'' corresponds to weakly relativistic configurations. It is a
relativistic generalization of the results obtained by
Antonov \cite{antonov}, Lynden-Bell and Wood \cite{lbw} and Katz
\cite{katzpoincare1} for the nonrelativistic classical gas.
Indeed, in the
nonrelativistic limit $\nu\rightarrow 0$, if we use the normalized variables
$\Lambda=-ER/GN^2m^2$ and $\eta=GNm^2/R k_B
T_{\infty}$, the hot spiral is rejected at infinity and we recover the results
of \cite{antonov,lbw,katzpoincare1}. The caloric
curve $\eta(\Lambda)$ has a spiral (snail-like) structure (see
Fig. 3 of
\cite{katzpoincare1}). In the microcanonical ensemble the system
undergoes a gravothermal
catastrophe below a minimum energy 
$E_{c}=-0.335\, GM^2/R$, corresponding to a density contrast ${\cal
R}_{\rm MCE}=\rho(0)/\rho(R)=709$.  This leads to a binary star
surrounded by a hot halo
\cite{inagakilb}. In the canonical ensemble, it undergoes
an
isothermal collapse below a
minimum temperature $T_{c}=GMm/(2.52Rk_B)$, corresponding to a density
contrast ${\cal
R}_{\rm CE}=32.1$. This leads to a Dirac peak containing all the
particles \cite{post}.

The ``hot spiral''  corresponds to strongly relativistic
configurations. It is related
(but not identical) to the caloric curve of the
box-confined self-gravitating black-body
radiation obtained by Sorkin {\it et al.} \cite{sorkin} and Chavanis
\cite{aarelat2}, which also has the form of a spiral
(see Fig. 15 of
\cite{aarelat2}). For the self-gravitating black-body radiation the system
undergoes a gravitational
collapse (presumably towards a black hole) above a maximum energy ${M}_{\rm
max}c^2=0.246\, Rc^4/G$ in the microcanonical ensemble, corresponding to an
energy density contrast
${\cal
R}_{\rm MCE}=\epsilon(0)/\epsilon(R)=22.4$, or
above a
maximum temperature $k_B(T_\infty)_{\rm max}=0.445\,(\hbar^3c^7/GR^2)^{1/4}$ in
the canonical ensemble,
corresponding to an energy density contrast ${\cal
R}_{\rm CE}=\epsilon(0)/\epsilon(R)=1.91$. In
Sec. \ref{sec_url}
of this paper, we
discuss in detail
the connection between the hot spiral of the general relativistic classical 
gas
and of the self-gravitating black-body radiation. We show that for
$\nu\rightarrow
0$,   if we
use the normalized variables ${\cal M}=GM/Rc^2$ and 
${\cal B}=Rc^4/GNk_B T_{\infty}$, instead of $\Lambda$ and $\eta$, the cold
spiral  is rejected at infinity. In that case, the caloric curve of the general
relativistic classical self-gravitating gas
tends to a  limit curve ${\cal B}({\cal M})$ which has a spiral (snail-like)
structure. The hot spiral corresponds to ultrarelativistic configurations.
The maximum
mass $M_{\rm max}=0.246\, Rc^2/G$ and the corresponding energy
density contrast
$22.4$ are the same as
for the self-gravitating black-body radiation obtained in
\cite{sorkin,aarelat2} but the maximum temperature $k_B T_{\rm
max}=0.0561\, Rc^4/GN$
and the corresponding energy
density contrast $10.3$ are
different from the self-gravitating  black-body radiation because they have a
different physical origin.

The paper is organized as follows. In Sec.
\ref{sec_smnrf} we develop the statistical
mechanics of nonrelativistic self-gravitating classical particles.
In Sec. \ref{sec_grf} we develop
the statistical mechanics of general relativistic classical
particles. In Sec. \ref{sec_gc} we present a scaling argument showing that
the caloric curve of general relativistic  classical particles depends on a
unique control parameter $\nu=GNm/Rc^2$ (compactness parameter). Generically,
the caloric curve has the form of a double spiral. In Sec. \ref{sec_cold}
we consider the nonrelativistic
limit and show that, when $\nu\rightarrow 0$, the normalized caloric curve
$\eta(\Lambda)$ tends towards a limit curve exhibiting
a single cold spiral, as found in former
works \cite{antonov,lbw,katzpoincare1}. In Sec.
\ref{sec_url} we consider the ultrarelativistic
limit and show that, when $\nu\rightarrow 0$, the normalized caloric curve
${\cal B}({\cal M})$ tends towards a limit curve exhibiting a single hot
spiral. This asymptotic curve has not been reported previously. We
discuss the analogies and the differences
between this asymptotic caloric curve and the caloric curve of the
self-gravitating black-body radiation \cite{sorkin,aarelat2}.
Finally, in Appendices \ref{sec_scc} and \ref{sec_sccnrb} we compare
box-confined isothermal models with heavily
truncated isothermal distributions in Newtonian gravity and general
relativity.

\section{Statistical mechanics of nonrelativistic self-gravitating  classical
particles}
\label{sec_smnrf}

In this section, we consider the statistical mechanics of nonrelativistic
self-gravitating
 classical particles.
The formalism and
the notations are the same as in Sec. II of
Paper I in the case of fermions.\footnote{As recalled in the
Introduction, the formalism developed in Paper I is valid for an arbitrary form
of entropy.} We shall not repeat the equations that are identical. The
only difference is that
we are considering classical particles described by the Boltzmann entropy
density
\begin{eqnarray}
\label{a1}
s(r)&=&-k_B \int f\left\lbrack \ln\left (\frac{f}{f_{*}}\right
)-1\right\rbrack \, {d}{\bf p}.
\end{eqnarray}
The Boltzmann entropy
can be obtained from a combinatorial analysis. It
is equal to the logarithm of the number of microstates
(complexions) corresponding to a given
macrostate (see \cite{ijmpb} for details). A
{\it microstate} is
characterized by the specification of the position ${\bf r}_i$  and the impulse
${\bf p}_i$ of all the particles ($i=1,...,N$). A {\it macrostate} is
characterized by
the (smooth)
distribution function $f({\bf r},{\bf p})$ giving the density of particles in
the macrocell (${\bf r}, {\bf r}+d{\bf r}; {\bf p}, {\bf p}+d{\bf p})$,
irrespectively of their precise position in the cell. Since a classical system
is ``diluted'' in phase space, we do not have to put any
constraint
on the possible microstates. The
combinatorial analysis then directly leads to the Boltzmann
entropy (\ref{a1}) where $f_{*}$ is a constant introduced for dimensional
reasons (it is related to the size $h$ of a microcell). If
the
particles are fermions, as in Paper I, the Boltzmann entropy (\ref{a1}) can also
be
obtained by expanding the Fermi-Dirac entropy (I-22)\footnote{Here and in the
following (I-x) refers to Eq. (x) of Paper I.} for $f\ll f_{\rm max}$, where 
$f_{\rm max}=g/h^3$ is the maximum possible value of the distribution
function fixed by the Pauli exclusion principle ($h$ is the Planck constant and 
$g$ is the spin multiplicity of quantum
states). This establishes $f_*=f_{\rm max}$. More generally, all
the results of the present  paper (which are valid for classical particles) can 
be
obtained from the results of Paper I (which are valid for fermions) by
considering
the
nondegenerate limit $f\ll f_{\rm max}$. However, the present formalism is more
general since it can also describe a gas of bosons far from the condensation
point ($T\gg T_c$). We recall that a statistical
equilibrium
state exists only if the system is confined within a box of radius $R$ otherwise
it would evaporate (see, e.g., Ref. \cite{sc} for
details). We shall therefore consider the ``box model'' as in Paper I.

\subsection{Maximization of the entropy density at fixed kinetic energy
density and particle number density}
\label{sec_lmnr}

In the microcanonical ensemble, the statistical equilibrium state is obtained by
maximizing the entropy $S$ at fixed energy $E$ and particle
number $N$. As in Paper I, we proceed in two steps. We first maximize
the
Boltzmann entropy density $s(r)$ at fixed kinetic
energy 
density $\epsilon_{\rm kin}(r)$ and particle number density $n(r)$ with
respect to variations on $f({\bf r},{\bf
p})$. This  leads to the
Maxwell-Boltzmann distribution function
\begin{equation}
\label{a2}
f({\bf r},{\bf p})=\frac{g}{h^3}e^{-\beta(r)p^2/2m+\alpha(r)},
\end{equation}
where $\beta(r)$ and $\alpha(r)$ are local Lagrange multipliers. Introducing the
local temperature $T(r)$ and the local chemical potential $\mu(r)$ through the
relations
$\beta(r)=1/k_B T(r)$ and $\alpha(r)=\mu(r)/k_B T(r)$, the Maxwell-Boltzmann
distribution (\ref{a2}) can be rewritten as
\begin{equation}
\label{a3}
f({\bf r},{\bf p})=\frac{g}{h^3} e^{-\lbrack p^2/2m-\mu(r)\rbrack/k_B
T(r)}.
\end{equation}
It corresponds to the condition of local thermodynamical equilibrium.
Substituting the Maxwell-Boltzmann distribution (\ref{a3}) into Eqs.
(I-17), (I-18) and (I-20), and performing the Gaussian integrations, we get
\begin{equation}
\label{a4}
n(r)=\frac{g}{h^3}\lbrack 2\pi m k_B
T(r)\rbrack^{3/2}e^{\mu(r)/k_B T(r)},
\end{equation}
\begin{equation}
\label{a5}
\epsilon_{\rm kin}(r)=\frac{3}{2}n(r)k_B T(r),
\end{equation}
\begin{equation}
\label{a6}
P(r)=\frac{2}{3}\epsilon_{\rm kin}(r)=n(r)k_B T(r),
\end{equation}
where we recall that the first equality in Eq. (\ref{a6}) is valid for an
arbitrary
nonrelativistic perfect gas, whatever its distribution function (see
Appendix A of Paper I). On the other hand, the second equality of Eq.
(\ref{a6}) is the famous Boyle's law of a classical gas.
These equations relate the Lagrange
multipliers 
$\beta(r)$ and $\alpha(r)$, or the thermodynamical variables $T(r)$ and
$\mu(r)$,
to the
constraints $\epsilon_{\rm
kin}(r)$ and $n(r)$. Substituting the Maxwell-Boltzmann distribution (\ref{a3})
into
the entropy density (\ref{a1}) we obtain the integrated Gibbs-Duhem relation
(I-37). It is shown in Appendix E of Paper I that this relation is valid
for an arbitrary form of entropy. For the Boltzmann entropy, using Eqs.
(\ref{a5})
and (\ref{a6}), the integrated Gibbs-Duhem relation (I-37) reduces to the form
\begin{equation}
\label{a7}
\frac{s(r)}{k_B}=\frac{5}{2}n(r)-\alpha(r)n(r).
\end{equation}
Using Eq. (\ref{a4}) it can also be written as
\begin{equation}
\label{a8b}
\frac{s(r)}{k_B}=\frac{5}{2}n(r)-n(r)\ln n(r)-n(r)\ln \left (\frac{h^3}{g}\right
)+\frac{3}{2}n(r)\ln \lbrack 2\pi m k_B
T(r)\rbrack.
\end{equation}
This equation provides an expression of the entropy density for a classical
system of particles in local thermodynamic equilibrium.

\subsection{Maximization of the entropy at fixed 
energy and particle
number}
\label{sec_gmnr}

We now maximize the entropy $S$ at  at fixed
energy $E$ and particle
number $N$ with respect to variations on $n(r)$ and $\epsilon_{\rm kin}(r)$.
Here, we just consider the extremization problem (see Appendix
\ref{sec_zalt} for
more
general results). As shown in Paper I, it
leads to the relations
\begin{equation}
\label{a9}
\beta=\frac{1}{k_B T}\qquad {\rm and}\qquad \alpha(r)=\alpha_0-\beta m\Phi(r),
\end{equation}
where $\beta=1/(k_B T)$ and $\alpha_0$ are constant. Therefore, at statistical
equilibrium
the temperature
is uniform and the chemical potential is given by the Gibbs law
\begin{equation}
\label{a9f}
\mu(r)=\mu_0-m\Phi(r)
\end{equation}
with $\mu_0=\alpha_0 k_B T$. The Gibbs law expresses the fact that
the {\it total} chemical potential $\mu_{\rm tot}\equiv \mu(r)+m\Phi(r)$
is uniform at statistical equilibrium. Substituting these
relations into Eqs. (\ref{a2}) and (\ref{a3}), we
obtain
the mean field Maxwell-Boltzmann
distribution function
\begin{equation}
\label{a9b}
f({\bf r},{\bf p})=\frac{g}{h^3}e^{\alpha_0}e^{-\beta ({p}^2/2m+m\Phi(r))}.
\end{equation}
This result can also be directly obtained by extremizing the 
entropy $S$ at  at fixed energy $E$ and particle number $N$ with
respect to variations on $f({\bf r},{\bf p})$ as detailed in Appendix C of
Paper I.
The local variables become
\begin{equation}
\label{a10}
n(r)=\frac{g}{h^3}e^{\alpha_0}\left (\frac{2\pi m}{\beta}\right )^{3/2}e^{-\beta
m\Phi(r)},
\end{equation}
\begin{equation}
\label{a11}
\epsilon_{\rm kin}(r)=\frac{3}{2} n(r) k_B T,
\end{equation}
\begin{equation}
\label{a12}
P(r)=n(r) k_B T,
\end{equation}
\begin{equation}
\label{a13}
\frac{s(r)}{k_B}=\frac{5}{2}n(r)-\alpha_0 n(r)+\beta n(r)m \Phi(r),
\end{equation}
\begin{equation}
\label{a14}
\frac{s(r)}{k_B}=\frac{5}{2}n(r)-n(r)\ln n(r)-n(r)\ln \left (\frac{h^3}{g}\right
)+\frac{3}{2}n(r)\ln ( 2\pi m k_B
T).
\end{equation}
These equations characterize a barotropic gas with a linear equation of state
$P(r)=n(r) k_B T$. Using Eq. (\ref{a10}), the distribution function (\ref{a9b})
can be written in terms of the density as
\begin{equation}
\label{a17}
f({\bf r},{\bf p})=n({r}) \left (\frac{\beta}{2\pi m}\right
)^{3/2} e^{-\beta {p}^2/2m}.
\end{equation}
On the other hand, the energy (kinetic $+$ potential) and the entropy are given
by
\begin{equation}
\label{a15}
E=E_{\rm kin}+W=\frac{3}{2}N k_B T+\frac{1}{2}\int \rho(r)\Phi(r)\, d{V},
\end{equation}
\begin{equation}
\label{a16}
S=-\frac{\mu_0}{T}N+\frac{5}{2}Nk_B+\frac{2W}{T},
\end{equation}
where $\rho(r)=n(r)m$ is the mass density.
We recall that the condition of statistical equilibrium, corresponding
to the extremization of the entropy at fixed energy and particle number, implies
the condition of hydrostatic equilibrium (see Appendix D of Paper I). This
is valid for a general form of entropy. In the present case, this
can be checked immediately by taking the gradient of the pressure given by Eq.
(\ref{a12}) and using Eq. (\ref{a10}).

\subsection{Canonical ensemble: Minimization of the free energy at fixed
particle number}

In the previous sections, we considered the microcanonical ensemble.
In the canonical ensemble, the statistical equilibrium state is obtained by
minimizing the free energy 
\begin{equation}
\label{a21WQng}
F=E-TS
\end{equation}
at
fixed particle number $N$. Proceeding similarly to
Sec. II.F of Paper I, we get the
same results as in Secs. \ref{sec_lmnr} and \ref{sec_gmnr} (for the first
variations). At
statistical equilibrium,
using Eqs. (\ref{a15})
and (\ref{a16}), the free energy is given by
\begin{equation}
\label{a21}
F=\mu_0 N-W-Nk_B T.
\end{equation}

\section{Statistical mechanics of general relativistic classical particles}
\label{sec_grf}

In this section, we consider the statistical mechanics of  self-gravitating
classical particles within the framework of general relativity. The formalism
and the notations are the same as in Sec. III of Paper I
but
now
the entropy density is given by
the Boltzmann entropy density (\ref{a1}) instead of the Fermi-Dirac entropy
density (I-121). As in the Newtonian case, a statistical
equilibrium state exists only if the system is confined within a box of radius
$R$ otherwise
it would evaporate.

\subsection{Maximization of the entropy density at fixed energy density and
particle
number density}
\label{sec_step1gr}

In the microcanonical ensemble, the statistical equilibrium state is obtained by
maximizing the entropy $S$ at fixed mass-energy $Mc^2$ and
particle number $N$. As in Paper I, we proceed in two steps. We first 
maximize the entropy density at fixed energy density $\epsilon(r)$ and
particle
number density $n(r)$ with respect to variations on $f({\bf r},{\bf p})$. This
leads to
the relativistic Maxwell-Boltzmann (or Maxwell-Juttner) distribution
function
\begin{equation}
\label{a23}
f({\bf r},{\bf p})=\frac{g}{h^3} e^{-\beta(r) E(p)+\alpha(r)},
\end{equation}
where $\beta(r)$ and $\alpha(r)$ are local Lagrange multipliers. Introducing the
temperature and the chemical potential through the relations
$\beta(r)=1/k_B T(r)$ and $\alpha(r)=\mu(r)/k_B T(r)$, it can be rewritten as
\begin{equation}
\label{a24}
f({\bf r},{\bf p})=\frac{g}{h^3}e^{-\lbrack
E(p)-\mu(r)\rbrack/k_B T(r)}.
\end{equation}
This corresponds to the condition of local thermodynamical
equilibrium. The local variables are
\begin{equation}
\label{a25}
n(r)=\frac{g}{h^3}\int e^{-\lbrack
E(p)-\mu(r)\rbrack/{k_B T(r)}} \, d{\bf p},
\end{equation}
\begin{equation}
\label{a26}
\epsilon(r)=\frac{g}{h^3}\int e^{-\lbrack
E(p)-\mu(r)\rbrack/{k_B T(r)}}E \, d{\bf p},
\end{equation}
\begin{equation}
\label{a27}
\epsilon_{\rm kin}(r)=\frac{g}{h^3}\int e^{-\lbrack
E(p)-\mu(r)\rbrack/{k_B T(r)}}E_{\rm kin}(p) \, d{\bf p},
\end{equation}
\begin{equation}
\label{a28}
P(r)=\frac{g}{3h^3}\int e^{-\lbrack
E(p)-\mu(r)\rbrack/k_B T(r)} p\frac{{d}E}{{d}p}  \, {d}{\bf p}.
\end{equation}
Using an integration by parts in the last equation, we get\footnote{This result
is valid for an arbitrary function $E(p)$.}
\begin{equation}
\label{a29}
P(r)=n(r) k_B T(r).
\end{equation}
This is the same equation of state as for a nonrelativistic gas (Boyle's
law). This result was
first established by Juttner \cite{juttner1,juttner2} but is implicit in the
work of Planck
\cite{planck}. Eqs.
(\ref{a25}) and (\ref{a26})
determine the
Lagrange
multipliers $\beta(r)$ and $\alpha(r)$, or the thermodynamical variables $T(r)$
and $\mu(r)$,
in terms
of $\epsilon(r)$ and $n(r)$. Substituting the Maxwell-Juttner
distribution function (\ref{a24}) into Eq. (\ref{a1}), and using Eq. 
(\ref{a25})-(\ref{a29}), we obtain the integrated Gibbs-Duhem
relation (I-135). It is shown in Appendix E of Paper I that this
relation is valid for an arbitrary form of entropy. For the Boltzmann entropy,
using Eq. (\ref{a29}), the integrated Gibbs-Duhem relation (I-135) reduces to
the form
\begin{equation}
\label{a30}
s(r)=\frac{\epsilon(r)-\mu(r) n(r)}{T(r)}+k_B n(r).
\end{equation}

\subsection{Maximization of the entropy at fixed mass-energy and
particle
number}
\label{sec_step2gr}

We now maximize the entropy $S$ at  at fixed
energy $E$ and particle
number $N$ with respect to variations on $n(r)$ and $\epsilon(r)$.
Here, we just consider the extremization problem. As
shown in Paper I, it
leads to the relation
\begin{equation}
\label{a31}
\alpha=\frac{\mu(r)}{k_B T(r)}=\frac{\mu_{\infty}}{k_B T_{\infty}},
\end{equation}
where $\alpha$ is a constant ($\mu_{\infty}$ and $T_{\infty}$ are the chemical
potential and the temperature measured by an observer at infinity). The
extremization problem also yields the TOV
equations [see Eqs. (I-155) and (I-156)] expressing the
condition of hydrostatic equilibrium and the Tolman-Klein relations [see Eqs.
(I-158) and (I-159)]. We recall
that, in general relativity, the temperature $T(r)$ depends on the position
even at statistical equilibrium. This is the so-called Tolman effect
\cite{tolman}.
Using Eqs. (\ref{a30}) and (\ref{a31}), we find that the 
entropy is given at statistical equilibrium by
\begin{equation}
\label{sgr}
S=\int_0^R \frac{\epsilon(r)}{T(r)}\left \lbrack
1-\frac{2GM(r)}{rc^2}\right
\rbrack^{-1/2}4\pi
r^2\, dr+k_B N-\frac{\mu_{\infty}}{T_{\infty}}N.
\end{equation}

\subsection{Canonical ensemble: Minimization of the free energy at fixed
particle number}
\label{sec_fece}

In the previous sections, we considered the microcanonical ensemble.
In the canonical ensemble, the statistical equilibrium state is obtained by
minimizing the free energy 
\begin{equation}
\label{a21WQ}
F=E-T_{\infty}S
\end{equation}
at fixed particle number $N$. Proceeding similarly to
Sec. III.F of Paper I, we get the
same results as in Secs. \ref{sec_step1gr} and \ref{sec_step2gr}. At
statistical equilibrium, the free energy is given by
\begin{eqnarray}
\label{b119}
F=(M-Nm)c^2
-\int_0^R \frac{T_{\infty}}{T(r)}\epsilon(r)\left \lbrack
1-\frac{2GM(r)}{rc^2}\right
\rbrack^{-1/2}4\pi
r^2\, dr-k_B T_{\infty}N+\mu_{\infty} N.
\end{eqnarray}

\subsection{Equations determining the statistical equilibrium state in terms
of $T(r)$}
\label{sec_eht}

\subsubsection{Local variables in terms of $T(r)$ and $\alpha$}

Using Eq. (\ref{a31}), we can rewrite the distribution function (\ref{a23})
or (\ref{a24}) and the local variables
(\ref{a25})-(\ref{a28}) as 
\begin{equation}
\label{a32}
f({\bf r},{\bf p})=\frac{g}{h^3}e^{\alpha}e^{-E(p)/k_B T(r)}, 
\end{equation}
\begin{equation}
\label{a33}
n(r)=\frac{g}{h^3}e^{\alpha}\int e^{-E(p)/k_B T(r)}  \, d{\bf p},
\end{equation}
\begin{equation}
\label{a34}
\epsilon(r)=\frac{g}{h^3}e^{\alpha}\int e^{-E(p)/k_B T(r)}  E \, d{\bf p},
\end{equation}
\begin{equation}
\label{a35}
\epsilon_{\rm kin}(r)=\frac{g}{h^3}e^{\alpha}\int e^{-E(p)/k_B T(r)} E_{\rm
kin}(p)
\,
d{\bf p},
\end{equation}
where we recall that $\alpha$ is a constant while the temperature $T(r)$ depends
on the position (Tolman's effect). The pressure is given by Eq. (\ref{a29}).

\subsubsection{Juttner transformation}
\label{sec_juttner}

In special relativity, the energy of a particle is
$E(p)=\sqrt{p^2c^2+m^2c^4}$. Making the
Juttner transformation \cite{juttner1,juttner2,chandrabook} 
\begin{equation}
\label{a36}
\frac{p}{mc}=\sinh\theta,
\end{equation}
we find that $E=mc^2\cosh\theta$. We also introduce  the normalized local
inverse
temperature
\begin{equation}
\label{a37}
b(r)=\frac{mc^2}{k_B T(r)}.
\end{equation}
With the transformation (\ref{a36}), Eqs.
(\ref{a33})-(\ref{a35}) become
\begin{equation}
\label{a38}
n(r)=\frac{4\pi g m^3c^3}{h^3}e^{\alpha}\int_0^{+\infty} e^{-b(r)\cosh\theta} 
\sinh^2\theta \cosh\theta\, d\theta,
\end{equation}
\begin{equation}
\label{a39}
\epsilon(r)=\frac{4\pi g m^4c^5}{h^3}e^{\alpha}\int_0^{+\infty}
e^{-b(r)\cosh\theta} 
\sinh^2\theta \cosh^2\theta\, d\theta,
\end{equation}
\begin{eqnarray}
\label{a40}
\epsilon_{\rm kin}(r)=\frac{4\pi g m^4c^5}{h^3}e^{\alpha}\int_0^{+\infty}
e^{-b(r)\cosh\theta}\sinh^2\theta \cosh\theta(\cosh\theta-1)\, d\theta.
\end{eqnarray}
They can be expressed in terms of the  modified Bessel functions defined by
\begin{equation}
\label{a41}
K_n(z)=\int_0^{+\infty}e^{-z\cosh\theta}\cosh(n\theta)\, d\theta.
\end{equation}
For future reference, we recall their asymptotic behaviors: 
\begin{equation}
\label{lu1}
K_n(z)\sim \frac{1}{2}\frac{(n-1)!}{\left (\frac{1}{2}z\right )^n}
\qquad (z\rightarrow 0),
\end{equation}
\begin{equation}
\label{a65}
K_n(z)\simeq \left (\frac{\pi}{2z}\right )^{1/2}e^{-z}\left
(1+\frac{4n^2-1}{8z}+...\right )\qquad (z\rightarrow +\infty).
\end{equation}
The nonrelativistic limit ($k_B T\ll mc^2$) corresponds to $z\rightarrow
+\infty$ and
the ultrarelativistic limit ($k_B T\gg mc^2$) corresponds to $z\rightarrow 0$.
Using the relations
\begin{equation}
\label{a42}
\sinh^2\theta\cosh\theta=\frac{1}{4}\lbrack \cosh(3\theta)-\cosh\theta\rbrack,
\end{equation}
\begin{equation}
\label{a43}
\sinh^2\theta\cosh^2\theta=\frac{1}{8}\lbrack \cosh(4\theta)-1\rbrack,
\end{equation}
and  the recurrence formula
\begin{equation}
\label{a44}
K_{n-1}(z)-K_{n+1}(z)=-\frac{2n}{z}K_n(z),
\end{equation}
we can write the local variables as
\begin{equation}
\label{a45}
n(r)=\frac{4\pi gm^3c^3}{h^3}e^{\alpha}\frac{1}{b(r)}K_2(b(r)),
\end{equation}
\begin{equation}
\label{a46}
\epsilon(r)=\frac{4\pi gm^4c^5}{h^3}e^{\alpha}\frac{1}{b(r)}K_2(b(r))\left
\lbrack
\frac{K_1(b(r))}{K_2(b(r))}+\frac{3}{b(r)}\right \rbrack,
\end{equation} 
\begin{equation}
\label{a47}
\epsilon_{\rm kin}(r)=\frac{4\pi
gm^4c^5}{h^3}e^{\alpha}\frac{1}{b(r)}K_2(b(r))\left
\lbrack
\frac{K_1(b(r))}{K_2(b(r))}+\frac{3}{b(r)}-1\right \rbrack,
\end{equation} 
\begin{equation}
\label{a48}
P(r)=n(r)\frac{mc^2}{b(r)}.
\end{equation}
On the other hand, using Eq. (\ref{a45}), the distribution function (\ref{a32})
can
be written in terms  of the density as
\begin{equation}
\label{a48b}
f({\bf r},{\bf
p})=\frac{n(r)}{4\pi m^3c^3}\frac{b(r)}{K_2(b(r))}e^{-b(r)E(p)/mc^2}.
\end{equation}
This expression, which can be compared to Eq. (\ref{a17}) in the nonrelativistic
case, is tricky because the factor in
front of the exponential is independent of $r$ [see Eq. (\ref{a32})]. The
spatial
inhomogeneity of the system due to the self-gravity manifests itself in the
temperature $T(r)$ as discussed in Paper I.

\subsubsection{Equation of state}
\label{sec_eos}

From Eqs.
(\ref{a45})-(\ref{a47}) we can write
\begin{equation}
\epsilon_{\rm kin}(r)=n(r)mc^2{\cal F}(b(r)),\qquad
\epsilon(r)=n(r)mc^2\left\lbrack 1+{\cal F}(b(r))\right\rbrack,
\label{sandr}
\end{equation}
where we have introduced  the function \cite{aarelat1}:
\begin{equation}
{\cal F}(z)=\frac{K_1(z)}{K_2(z)}+\frac{3}{z}-1.
\end{equation} 
Its asymptotic behaviors are 
\begin{equation}
\label{ff1}
{\cal F}(z)\sim \frac{3}{z}
\qquad (z\rightarrow 0),
\end{equation}
\begin{equation}
\label{ff2}
{\cal F}(z)\sim \frac{3}{2z} \qquad (z\rightarrow +\infty).
\end{equation}
Using Eqs. (\ref{a48}) and (\ref{sandr}), we get
\begin{equation} 
P(r)=\frac{\epsilon(r)}{b(r)\lbrack
1+{\cal F}(b(r))\rbrack}.
\end{equation} 
Since $b(r)$ is related to $\epsilon(r)$ through Eq.  (\ref{a46}), the foregoing
equation can be viewed as an equation of state relating the pressure to the
energy density. It is important to note, however, that this relation also
depends on $\alpha$, so that it is of the form $P(r)=P(\epsilon(r),\alpha)$.

The ratio between the pressure and the kinetic energy density is
\begin{equation}
\frac{P(r)}{\epsilon_{\rm kin}(r)}=\frac{1}{b(r){\cal F}(b(r))}.
\end{equation} 
In the nonrelativistic limit $b\rightarrow +\infty$ ($k_B T\ll mc^2$), using Eq.
(\ref{ff2}), we
get
\begin{equation}
\frac{P(r)}{\epsilon_{\rm kin}(r)}\rightarrow \frac{2}{3}.
\end{equation} 
In the ultrarelativistic limit $b\rightarrow 0$ ($k_B T\gg mc^2$), using Eq.
(\ref{ff1}), we get
\begin{equation}
\frac{P(r)}{\epsilon_{\rm kin}(r)}\rightarrow \frac{1}{3}.
\end{equation} 
This returns the general results from Appendix A of Paper I which are valid for
an arbitrary distribution function.

\subsubsection{The TOV equations in terms of $T(r)$}
\label{sec_tovt}

Using the general results of Paper I, the TOV
equations can be written in terms of $T(r)$ as
\begin{equation}
\label{b126}
\frac{dM}{dr}=\frac{\epsilon}{c^2}4\pi r^2,
\end{equation}
\begin{equation}
\label{b125}
\frac{1}{T}\frac{{d}T}{{d}r}=-\frac{1}{c^2}\frac{\frac{GM(r)}{r^2}+\frac{4\pi
G}{c^2}Pr}{1-\frac{2GM(r)}{r c^2}},
\end{equation}
with the boundary conditions
\begin{equation}
\label{b128}
M(0)=0\qquad {\rm and}\qquad T(0)=T_0.
\end{equation}
For a given value of $\alpha$ and $T_0$ one can solve Eqs.
(\ref{b126}) and (\ref{b125}) between $r=0$ and $r=R$ with the local variables
given by Eqs.
(\ref{a45})-(\ref{a48}). The
particle number
constraint
\begin{equation}
\label{b127}
N=\int_0^R n(r)\left \lbrack 1-\frac{2GM(r)}{rc^2}\right
\rbrack^{-1/2}4\pi
r^2\, dr
\end{equation}
can be used to determine $T_0$ as a function of  $\alpha$ (there may be several
solutions for the same value of $\alpha$). The total mass $M$ and the
temperature $T_{\infty}$ measured by an observer at infinity  are then obtained
from the relations
\begin{equation}
\label{b130}
M=M(R)\qquad {\rm and}\qquad T_{\infty}=T(R) \sqrt{1-\frac{2GM}{Rc^2}}.
\end{equation}
In this manner, we get the binding energy $E=(M-Nm)c^2$ and the Tolman
(global) temperature $T_{\infty}$ as a function of
$\alpha$.
By
varying $\alpha$ between $-\infty$ and $+\infty$, we can obtain the full caloric
curve $T_{\infty}(E)$ for a given value of $N$ and $R$ (we show in Sec.
\ref{sec_gcs} that the results depend only on the ratio $N/R$). Finally, the
entropy
and the free energy are given by Eqs. (\ref{sgr}) and
(\ref{b119}).

\subsubsection{The TOV equations in terms of $\varphi(r)$}

Introducing the  gravitational potential
$\varphi(r)$
through the relation (see Paper I): 
\begin{equation}
\label{a64}
\frac{k_B
T(r)}{mc^2}=\frac{1}{b(r)}=\frac{1}{|\alpha|}\sqrt{1+\frac{\varphi(r)}{c^2}},
\end{equation}
we can rewrite the local variables (\ref{a32})-(\ref{a35}) in terms of $\alpha$
and $\varphi(r)$. We can also
directly substitute the relation (\ref{a64}) into the Juttner
equations (\ref{a45})-(\ref{a48}). On the other hand, the TOV equations can be
written in terms of
$\varphi(r)$ as (see Paper I):
\begin{equation}
\label{b147}
\frac{dM}{dr}=\frac{\epsilon}{c^2}4\pi r^2,
\end{equation}
\begin{equation}
\label{b145}
\frac{d\varphi}{dr}=-\frac{2G}{c^2}\left \lbrack \frac{\varphi(r)}{c^2}+1\right
\rbrack \frac{M(r)c^2+4\pi P(r)r^3}{r^2
\left \lbrack 1-\frac{2GM(r)}{rc^2}\right \rbrack},
\end{equation}
with the boundary conditions
\begin{equation}
\label{b128b}
M(0)=0\qquad {\rm and}\qquad \varphi(0)=\varphi_0>-c^2.
\end{equation}
For a given value of $\alpha$ and $\varphi_0$ we can solve Eqs.
(\ref{b147}) and (\ref{b145}) between $r=0$ and $r=R$ with the local
variables obtained from Eqs.
(\ref{a45})-(\ref{a48}) and (\ref{a64}).
The particle number
constraint
\begin{equation}
\label{b127b}
N=\int_0^R n(r)\left \lbrack 1-\frac{2GM(r)}{rc^2}\right
\rbrack^{-1/2}4\pi
r^2\, dr
\end{equation}
can be used to determine $\varphi_0$  as a function of $\alpha$ (there may be
several solutions for the same value of $\alpha$). The total mass $M$ and the
temperature $T_{\infty}$ measured by an observer at infinity  are then obtained
from the
relations
\begin{equation}
\label{b130b}
M=M(R)\qquad {\rm and}\qquad
\frac{k_B T_{\infty}}{mc^2}=\frac{1}{|\alpha|}\sqrt{\frac{
\varphi(R)}{c^2}+1}\left
(1-\frac{2GM}{Rc^2}\right )^{1/2}.
\end{equation}
In this manner, we get the binding energy $E=(M-Nm)c^2$ and the Tolman (global)
temperature  $T_{\infty}$ as a function of
$\alpha$. By
varying $\alpha$ between $-\infty$ and $+\infty$, we can obtain the full caloric
curve $T_{\infty}(E)$ for a given value of $N$ and $R$ (we show in Sec.
\ref{sec_gcs} that the results depend only on the ratio $N/R$). Finally, the
entropy
and the free energy are given by Eqs. (\ref{sgr}) and
(\ref{b119}) with Eq. (\ref{a64}).

\section{General case: The double spiral}
\label{sec_gc}

In this section, we discuss general properties of the caloric curve of general
relativistic classical particles. A more detailed discussion is given in
\cite{roupas,acb}.

\subsection{The $N/R$ scaling}
\label{sec_gcs}

We first show that, for classical particles, the normalized
caloric
curve determined by the equations of Sec. \ref{sec_eht} depends only on the
ratio $N/R$
instead of $N$ and $R$ individually as in the case of fermions (Paper I). If
we introduce the scaled variables
\begin{equation}
\label{a49}
\tilde r=\frac{r}{R},\qquad \tilde M(\tilde r)=\frac{GM(r)}{Rc^2},\qquad
\tilde\epsilon(\tilde r)=\frac{GR^2\epsilon(r)}{c^4},\qquad
\tilde P(\tilde r)=\frac{GR^2 P(r)}{c^4},
\end{equation}
\begin{equation}
\label{a50}
\tilde n(\tilde
r)=\frac{GR^2 n(r)m}{c^2}, \qquad \tilde N(\tilde
r)=\frac{GN(r)m}{Rc^2},\qquad \tilde \alpha=\alpha+\ln \left (\frac{G m^4 c
R^2}{h^3}\right )
\end{equation}
in the equations of Sec. \ref{sec_eht}, we find that the local variables
(\ref{a45})-(\ref{a48}) become
\begin{equation}
\label{a51}
\tilde n(\tilde r)=4\pi g e^{\tilde\alpha}\frac{1}{b(\tilde r)}K_2(b(\tilde r)),
\end{equation}
\begin{equation}
\label{a52}
\tilde\epsilon(\tilde r)=4\pi g
e^{\tilde\alpha}\frac{1}{b(\tilde r)}K_2(b(\tilde r))\left
\lbrack
\frac{K_1(b(\tilde r))}{K_2(b(\tilde r))}+\frac{3}{b(\tilde r)}\right \rbrack,
\end{equation}
\begin{equation}
\label{a53}
\tilde P(\tilde r)=\tilde n(\tilde r)\frac{1}{b(\tilde r)}.
\end{equation}
On the other hand,  the TOV equations (\ref{b126}) and (\ref{b125})  become
\begin{equation}
\label{a55}
\frac{d\tilde M}{d\tilde r}=4\pi \tilde\epsilon(\tilde r){\tilde r}^2,
\end{equation}
\begin{equation}
\label{a54}
\frac{db}{d\tilde r}=b(\tilde r)\frac{\tilde M(\tilde r)+4\pi \tilde P(\tilde
r){\tilde r}^3}{{\tilde r}^2
\left \lbrack 1-\frac{2\tilde M(\tilde r)}{\tilde r}\right \rbrack},
\end{equation}
with the boundary conditions 
\begin{equation}
\label{a56}
 \tilde M(0)=0,\qquad b(0)=b_0\ge 0.
\end{equation}
The particle number constraint (\ref{b127}) takes the form
\begin{equation}
\label{a57}
\tilde N=\int_0^1 \tilde n(\tilde r)\left \lbrack
1-\frac{2G\tilde M(\tilde r)}{\tilde rc^2}\right \rbrack^{-1/2}4\pi
{\tilde r}^2\, {d}\tilde r.
\end{equation}
Finally, the mass and the inverse Tolman temperature $\beta_{\infty}=1/k_B
T_{\infty}$ [see Eq. (\ref{b130})] 
are given by
\begin{equation}
\label{a58}
\tilde M=\tilde M(1),\qquad b_{\infty}=b(1)\left (1-2\tilde M\right )^{-1/2}.
\end{equation}
In this manner, we see that
the problem depends only on the dimensionless number
\begin{equation}
\label{a59}
\nu=\frac{GNm}{Rc^2},
\end{equation}
corresponding to $\tilde N$. This is the so-called compactness parameter. It can
be interpreted as the ratio $\nu=R_S^*/R$
between the effective Schwarzschild radius $R_S^*=GNm/c^2$, defined in terms of
the rest mass $Nm$, and the box radius $R$. Alternatively, it can
be interpreted as the ratio $\nu=Nm/M_S^*$ between the rest mass $Nm$ and the
effective Schwarzschild mass $M_S^*=Rc^2/G$  defined in terms of
the box radius.   For a given
value of
$\nu$ 
(i.e. $\tilde N$), we
can solve the dimensionless equations (\ref{a51})-(\ref{a58}) and determine
\begin{equation}
\label{a60}
{\cal M}=\frac{GM}{Rc^2}\qquad {\rm and}\qquad b_{\infty}=\frac{mc^2}{k_B
T_{\infty}},
\end{equation}
where ${\cal M}$ corresponds to $\tilde M$. As a result, for
a given
value of $\nu$, we can plot the caloric curve giving $b_{\infty}$ as
a function of ${\cal
M}$. Actually, in order to make the link with the
nonrelativistic limit $c\rightarrow +\infty$, it
is preferable to plot
\begin{equation}
\label{a62}
\eta=\frac{\beta_{\infty}GNm^2}{R}=\frac{GNm^2}{k_B
T_{\infty}R}
\end{equation}
as a function of 
\begin{equation}
\label{a63}
\Lambda=-\frac{ER}{GN^2m^2}=-\frac{(M-Nm)c^2R}{GN^2m^2},
\end{equation} 
where $E=(M-Nm)c^2$ is the binding energy (see Paper I). We have the
relations
\begin{equation}
\label{a63b}
\Lambda=-\frac{{\cal M}-\nu}{\nu^2},\qquad \eta=b_{\infty}\nu.
\end{equation} 
In terms
of the variables $\eta$ and $\Lambda$, the nonrelativistic
caloric curve is recovered in the limit $\nu\rightarrow 0$ (see Sec.
\ref{sec_cold}).

{\it Remark:} The scaling of this section  amounts to taking $m=c=G=R=h=1$ in
the original equations. In that case, the normalized caloric curve is obtained
by plotting $\eta=\beta_{\infty}N$ as a function of $\Lambda=-E/N^2$ for a given
value of $N$ \cite{acb}. We note, however, that the entropy introduces a new
dependence in $R$. Indeed, the entropy and free energy scale as
\begin{equation}
\label{rw1}
S=\frac{Rc^2}{Gm}k_B \tilde S-N k_B \ln\left (\frac{Gm^4cR^2}{h^3}\right ),
\end{equation} 
\begin{equation}
\label{rw2}
F=\frac{Rc^4}{G} \tilde F+N k_B T_{\infty} \ln\left (\frac{Gm^4cR^2}{h^3}\right
),
\end{equation} 
where
\begin{equation}
\label{rw3}
\tilde S=\int_0^1 \tilde \epsilon(\tilde r)b(\tilde r)\left\lbrack
1-\frac{2\tilde M(\tilde r)}{\tilde r}\right\rbrack^{-1/2}4\pi {\tilde r}^2\,
d\tilde r+\tilde N-\tilde\alpha\tilde N,
\end{equation} 
and
\begin{equation}
\label{rw4}
\tilde F=\tilde M-\tilde N-\frac{\tilde S}{b_{\infty}}.
\end{equation} 
We see on Eq. (\ref{rw1}) that the entropy involves a contribution  $-N k_B \ln
(R^2)$  scaling like the logarithm of the area of the system. However, this
term appears just as an additive constant in the entropy so it can be omitted
in most applications.

\subsection{Description of the caloric curve}

\begin{figure}
\begin{center}
\includegraphics[clip,scale=0.3]{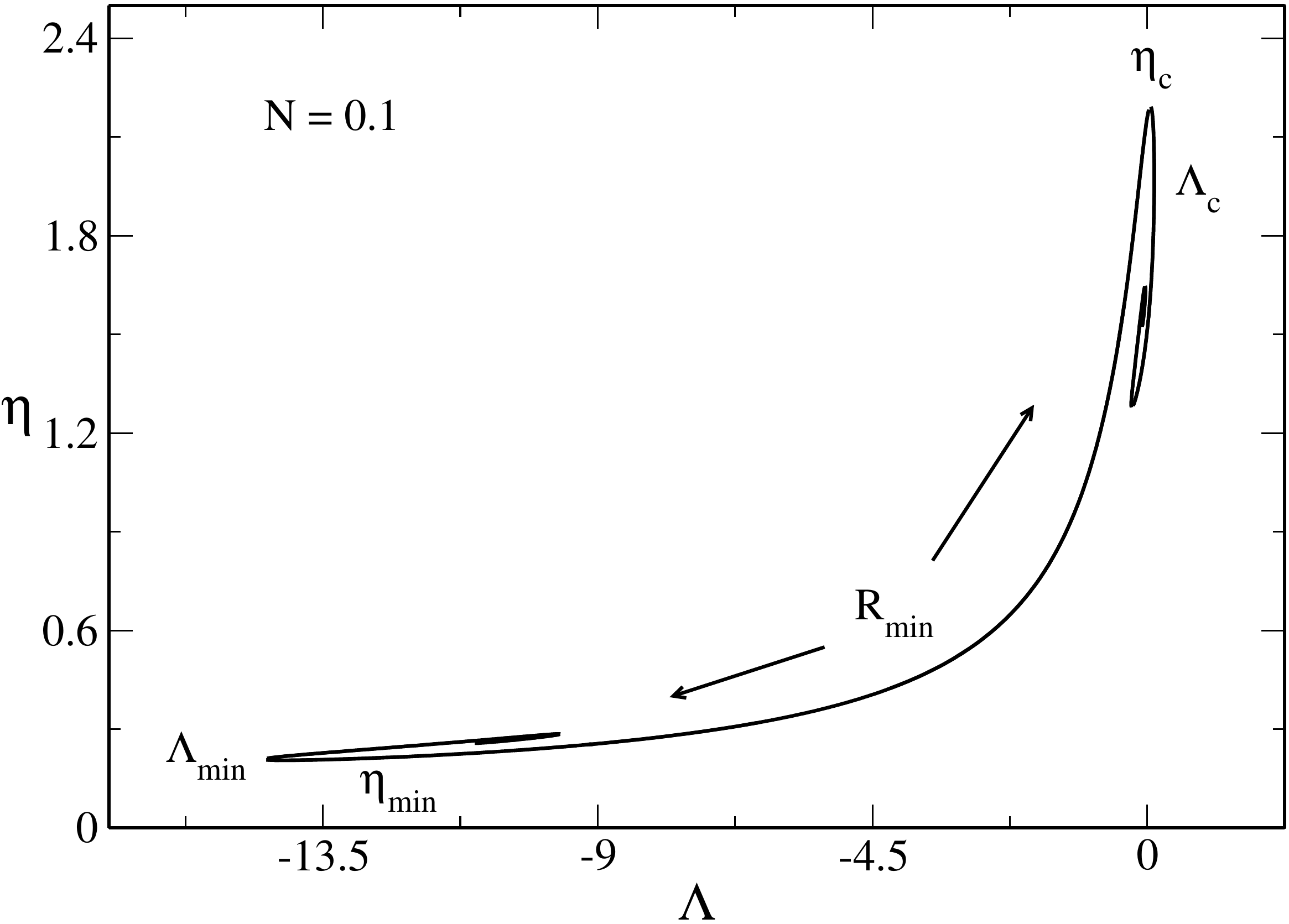}
\caption{Caloric
curve of a general relativistic  self-gravitating classical
gas for $\nu=0.1$. It presents a double spiral. The
system collapses at
low energies and low temperatures as in the case of a nonrelativistic 
self-gravitating classical gas (cold spiral). It also collapses
(presumably towards a
black hole) at high
energies and high temperatures as in the case of the
self-gravitating black-body radiation (hot spiral).}
\label{kcal_N01_linked_colorsPH}
\end{center}
\end{figure}

In Fig. \ref{kcal_N01_linked_colorsPH} we have plotted the
caloric curve of a
general relativistic classical self-gravitating gas for a typical value of
$\nu$ (specifically $\nu=0.1$). It has the form of a
double spiral parametrized by the energy density contrast ${\cal
R}=\epsilon(0)/\epsilon(R)$ \cite{roupas,acb}. The density contrast ${\cal R}$
is minimum at the
``center'' of the caloric curve and increases along the series of equilibria in
the two directions as we approach the spirals.

The cold spiral (on the right) is a relativistic generalization of the caloric
curve obtained by Katz \cite{katzpoincare1} for a
nonrelativistic self-gravitating classical gas in Newtonian
gravity (see Sec. \ref{sec_cold}). It corresponds to weakly  relativistic
configurations (except when
$\nu$ is large). It exhibits a minimum energy (in the microcanonical ensemble)
at
$E_c$ and a minimum temperature (in the canonical ensemble) at $T_c$ below
which the system
undergoes a gravitational collapse as in the case of nonrelativistic
systems. If the system is sufficiently relativistic
($\nu$ large), the collapse may lead to the formation of a black
hole. On the other hand, the hot spiral is a purely general
relativity result.
It is similar (but not identical) to the caloric curve
obtained by Chavanis \cite{aarelat2} for the self-gravitating black-body
radiation (see
Sec. \ref{sec_url}). It corresponds to strongly relativistic
configurations. It exhibits a
maximum energy (in the microcanonical ensemble) at $E_{\rm max}$ and a maximum
temperature (in the canonical ensemble) at
$T_{\rm max}$ above which the system undergoes a gravitational collapse leading
presumably to the formation of a black hole.

In the canonical ensemble the series of equilibria is stable on the main branch 
between $\eta_{\rm
min}$ and
$\eta_{c}$. According to the Poincar\'e criterion \cite{poincare}, it becomes
unstable
at the first 
turning points of temperature when the specific heat becomes infinite, passing
from positive to negative values. A new  mode of instability is lost at each
subsequent turning
point of
temperature as the spirals rotate clockwise. In the microcanonical ensemble the
series of equilibria is
stable on the main branch between $\Lambda_{\rm min}$ and
$\Lambda_{c}$. According to the Poincar\'e criterion \cite{poincare}, it becomes
unstable at the first
turning points of energy when the specific heat vanishes, passing from negative
to
positive values. A new  mode of instability is lost at each subsequent turning
point of
energy as the spirals rotate clockwise. There are two regions of ensembles
inequivalence (one on each spiral), between the turning points of
temperature and energy, i.e., in the first regions of negative specific heat.

It has to be noted that the stable equilibrium states are in fact metastable,
as there is no global maximum of entropy or global minimum of free energy for
classical self-gravitating systems \cite{antonov,lbw}. However, these metastable
states have tremendously long lifetimes, scaling as $e^N$, so they are stable
in practice \cite{lifetime}.

The
evolution of the caloric curve with $\nu$ is
described in detail in \cite{roupas,acb}. As $\nu$ increases, the cold and hot
spirals
approach each other, merge at $\nu'_S=0.128$, form a loop above
$\nu_S=0.1415$, reduce to a point at $\nu_{\rm max}=0.1764$,
and finally
disappear. The limit $\nu\rightarrow 0$ is discussed in
detail in the following sections.

According to Ipser's conjecture \cite{ipser80} (see Paper I and footnote 5),
dynamical and
thermodynamical
stability coincide in general relativity. However, as discussed in Paper I
and footnote 5,
we
expect that the growth rate of the dynamical instability is small for weakly
relativistic configurations (cold spiral) and large for strongly relativistic
configurations (hot spiral). In other words, the collapse at $E_c$ is
essentially a thermodynamical instability that takes place on a secular
timescale while the collapse at $E_{\rm max}$ is essentially a dynamical
instability taking place on a short timescale.

\section{Nonrelativistic limit: the cold spiral}
\label{sec_cold}

The nonrelativistic limit of the classical
isothermal gas corresponds to $k_B T\ll
mc^2$. It can be formally obtained by taking the
limits $T\rightarrow 0$,
$b\rightarrow +\infty$, or $c\rightarrow +\infty$ in the equations of Sec.
\ref{sec_grf}. Proceeding as in Sec. IV of Paper I, we
recover the results of Sec. \ref{sec_smnrf}.

\subsection{Differential equation for $\rho(r)$}

The equation of state of a nonrelativistic classical gas at
statistical equilibrium is given by [see Eq. (\ref{a12})]:
\begin{equation}
\label{a66}
P(r)=\rho(r)\frac{k_B T}{m}.
\end{equation}
This is the classical isothermal equation of state \cite{chandrabook}.
Substituting this equation of state into the
fundamental equation of hydrostatic equilibrium (I-8), we obtain the following
differential equation 
\begin{equation}
\frac{1}{r^2}\frac{d}{dr}\left (r^2 \frac{d\ln\rho}{dr}\right )=-4\pi
G\beta m \rho
\label{a67}
\end{equation}
for the density profile $\rho(r)$.

\subsection{Differential equation for $\Phi(r)$}

At statistical equilibrium, the density of the system is related to the
gravitational
potential
by the Boltzmann distribution [see Eq. (\ref{a10})]:
\begin{equation}
\label{a68}
\rho(r)=m \frac{g}{h^3}e^{\alpha_0}\left (\frac{2\pi m}{\beta}\right
)^{3/2}e^{-\beta
m\Phi(r)}.
\end{equation}
Substituting this relation  into the Poisson equation (I-1), we obtain the
following differential equation
\begin{equation}
\label{a69}
\Delta\Phi=4\pi G m \frac{g}{h^3}e^{\alpha_0}\left (\frac{2\pi m}{\beta}\right
)^{3/2}e^{-\beta
m\Phi}
\end{equation}
for the gravitational potential $\Phi(r)$. This is the
so-called Boltzmann-Poisson equation. Once the gravitational potential
$\Phi(r)$ has
been determined by solving Eq. (\ref{a69}) the density profile is obtained from
Eq. (\ref{a68}). We note
that Eqs. (\ref{a67}) and (\ref{a69}) are equivalent (this is
shown in
Paper I in the general case).

\subsection{The Emden equation}

The density profile (\ref{a68}) can be written as
\begin{equation}
\rho(r)=\rho_{0}e^{-\beta m(\Phi(r)-\Phi_0)},
\label{a70}
\end{equation}
where $\rho_{0}$ is the central density and $\Phi_0$ is the central potential.
The Boltzmann-Poisson equation (\ref{a69}) then becomes
\begin{equation}
\Delta\Phi=4\pi G \rho_{0}e^{-\beta m(\Phi-\Phi_{0})}.
\label{a71}
\end{equation}
Introducing the dimensionless
variables
\begin{equation}
\rho=\rho_{0}e^{-\psi(\xi)},\qquad \psi=\beta m(\Phi-\Phi_{0}),\qquad \xi=(4\pi
G\beta m\rho_{0})^{1/2}r,
\label{a73}
\end{equation}
into Eq. (\ref{a71}), we obtain the Emden equation
\begin{equation}
\frac{1}{\xi^2}\frac{d}{d\xi}\left (\xi^2 \frac{d\psi}{d\xi}\right )=e^{-\psi},
\label{a74}
\end{equation}
with the boundary conditions $\psi(0)=\psi'(0)=0$. The same equation is
obtained by substituting the dimensionless
variables (\ref{a73}) into the equation of hydrostatic equilibrium  
(\ref{a67}) \cite{chandrabook}.

\subsection{Inverse temperature}

If we denote by $a$ the value of $\xi$ at the edge of the box,\footnote{This
quantity was denoted $\alpha$ in Ref. \cite{aaiso}. Here, we use the notation
$a$ instead of $\alpha$ to avoid confusion with the variable $\alpha=\mu/k_B T$
introduced
in Sec. \ref{sec_smnrf}.} we
have
\begin{equation}
a=(4\pi G\beta m\rho_{0})^{1/2}R\qquad {\rm and}\qquad
\xi=a\frac{r}{R}.
\label{a75}
\end{equation}
Introducing the inverse normalized temperature 
\begin{equation}
\eta=\frac{\beta GMm}{R},
\label{a76}
\end{equation}
and using Newton's law (I-3) applied at $r=R$, we
get
\begin{equation}
\eta=a\psi'(a).
\label{a77}
\end{equation}
The same result can be obtained by integrating
the Emden equation  (\ref{a74}) multiplied by $\xi^2$ between $\xi=0$ and
$\xi=a$.

\subsection{Energy}

For a nonrelativistic system, the virial theorem can be written as (see
Appendix B of Paper I)
\begin{equation}
2E_{\rm kin}+W=3P_{b}V,
\label{a78}
\end{equation}
where $P_b=P(R)$ is the pressure at the edge of the box and $V=(4/3)\pi R^3$ is
the volume of the box. The total
energy of the system is given by
\begin{equation}
E=E_{\rm kin}+W=-E_{\rm kin}+3P_{b}V=-\frac{3}{2}Nk_B T+\frac{4\pi
R^3\rho(R)k_B T}{m},
\label{a79}
\end{equation}
where we have used Eqs. (\ref{a12}) and (\ref{a15}).
Introducing the normalized energy 
\begin{equation}
\Lambda=-\frac{ER}{GM^2},
\label{a81}
\end{equation}
and using Eqs.  (\ref{a73}), (\ref{a75}) and (\ref{a77}), we obtain
\begin{equation}
\Lambda=\frac{3}{2a\psi'(a)}-\frac{e^{-\psi(a)}}{\psi'(a)^2}.
\label{a82}
\end{equation}

\subsection{Entropy and  free energy}

The entropy is given by Eq. (\ref{a16}).
Using
\begin{equation}
W=E-E_{\rm kin}=E-\frac{3}{2}Nk_B T,
\label{a84}
\end{equation}
we get
\begin{equation}
S=-\frac{\mu_{0}}{T} N+\frac{2E}{T}-\frac{1}{2}Nk_B.
\label{a85}
\end{equation}
On the other hand, applying Eq. (\ref{a10}) at $r=R$ and using Eqs. (\ref{a73}),
(\ref{a75}) and $\Phi(R)=-GM/R$
[see Eq. (I-5)]  we find that
\begin{equation}
\alpha_{0}=2\ln(a)+\frac{1}{2}\ln\eta-\psi(a)-\eta-\ln\mu+
\ln2-\frac{1}{2}\ln\pi,
\label{a87}
\end{equation}
where 
\begin{equation}
\mu=g\frac{m^4}{h^3}\sqrt{512\pi^4G^3MR^3}
\label{mu}
\end{equation}
is the so-called degeneracy parameter \cite{ijmpb}.\footnote{It should not be
confused with the chemical potential (the notation $\mu$ introduced in
\cite{ijmpb} is somewhat
unfortunate). The degeneracy parameter plays an important role for
self-gravitating fermions as it
determines the shape of their caloric curves \cite{ijmpb}. For classical
particles, or for fermions in the nondegenerate limit, it just
appears as an additive constant in the Boltzmann entropy (see Eq.
(\ref{a88})) so it can be omitted in most applications. We note that the
nondegenerate limit corresponds formally
to $\mu\rightarrow +\infty$.} Recalling that $\mu_0=\alpha_0 k_B T$
and substituting Eq.
(\ref{a87}) into Eq.
(\ref{a85}), we finally obtain 
\begin{equation}
\frac{S}{Nk_B}=-\frac{1}{2}
\ln\eta-2\ln(a)+\psi(a)+\eta-2\Lambda\eta+\ln\mu+\frac{1}{2}
\ln\pi-\ln2-\frac{1}{2}.
\label{a88}
\end{equation}
The free energy (\ref{a21WQng}) is then given by
\begin{equation}
\frac{FR}{GM^2}=\frac{1}{2\eta}
\ln\eta+\frac{2}{\eta}\ln(a)-\frac{1}{\eta}
\psi(a)-1+\Lambda-\frac{1}{\eta}\ln\mu-\frac { 1 } { 2\eta }
\ln\pi+\frac{1}{\eta}\ln2+\frac{1}{2\eta}.
\label{a88b}
\end{equation}

\subsection{The caloric curve $\eta(\Lambda)$}

\begin{figure}
\begin{center}
\includegraphics[clip,scale=0.3]{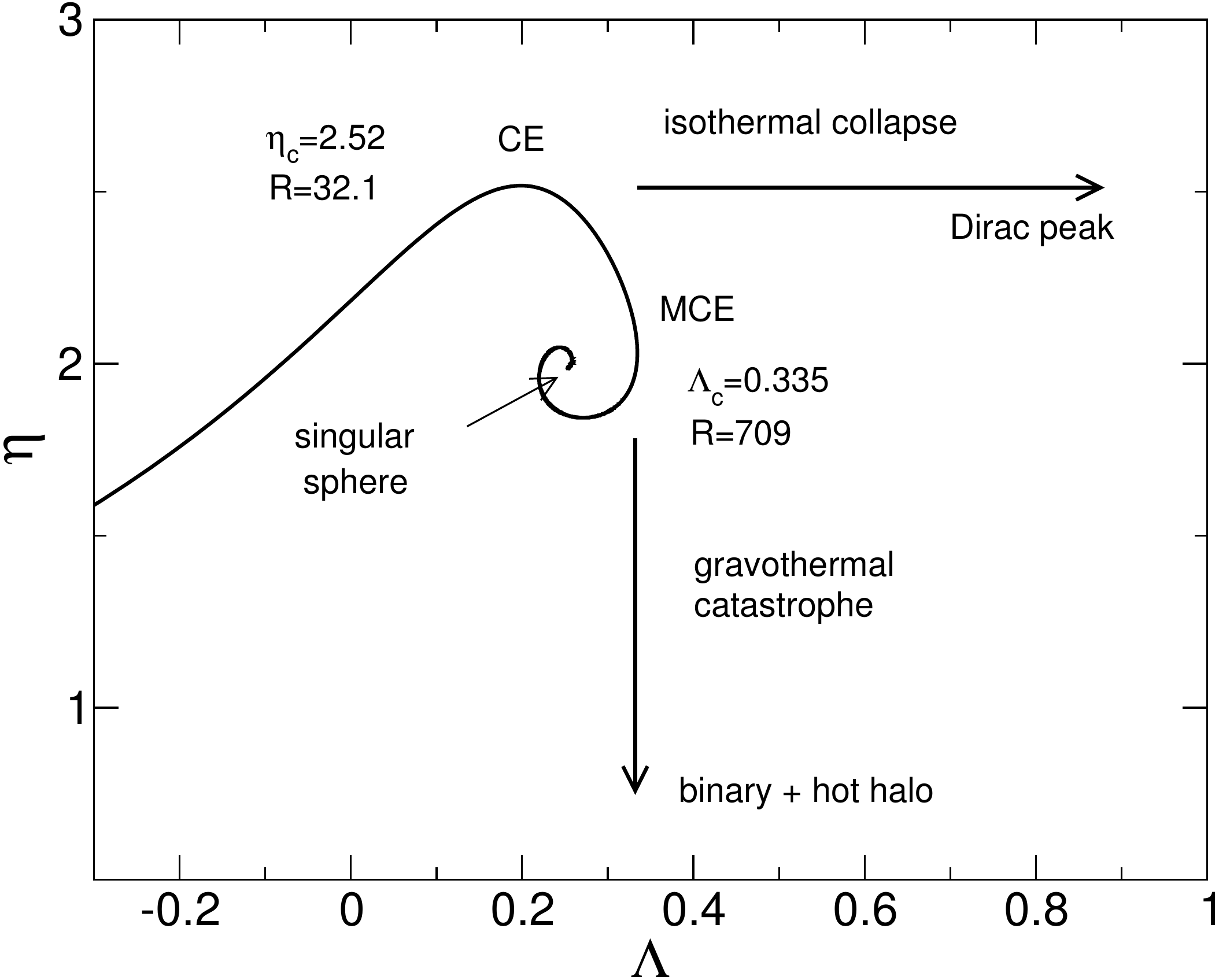}
\caption{Caloric curve $\eta(\Lambda)$  of the nonrelativistic 
self-gravitating classical
gas obtained by solving the Emden equation (\ref{a74}). It corresponds
to the limit curve of the general relativistic caloric
curve from Fig.
\ref{kcal_N01_linked_colorsPH} when $\nu\rightarrow 0$.}
\label{etalambda}
\end{center}
\end{figure}

The functions $\eta(a)$ and $\Lambda(a)$  defined
by Eqs. (\ref{a77}) and (\ref{a82}) can be
obtained by solving the Emden equation  (\ref{a74}) numerically.
They have been plotted in Ref. \cite{aaiso} and they display damped
oscillations.
From these
functions, using Eqs. (\ref{a77}) and (\ref{a82}), we can obtain the caloric
curve (or series of
equilibria) $\eta(\Lambda)$ parametrized by
$a$ or, equivalently, by the density contrast ${\cal
R}=\rho(0)/\rho(R)=e^{\psi(a)}$ \cite{aaiso}.  It has the form of a spiral (see
Fig.
\ref{etalambda}) along which the density contrast ${\cal R}$ increases. It
corresponds to
the cold spiral of the general relativistic caloric curve plotted in Fig.
\ref{kcal_N01_linked_colorsPH} in the nonrelativistic
limit $\nu\rightarrow 0$ and $\alpha\rightarrow +\infty$ (see
Sec. IV.C of Paper I and Ref. \cite{acb}). In
that limit, in which $k_B T\ll mc^2$, the hot spiral is rejected at infinity
($\eta_{\rm min}\rightarrow 0$ and $\Lambda_{\rm min}\rightarrow -\infty$)
when we use the variables $\eta$ and $\Lambda$.

The caloric curve exhibits a minimum energy (in the microcanonical ensemble) at
\begin{equation}
\label{lc}
\Lambda_{c}=0.335, \qquad a_c=34.4,\qquad {\cal R}_c=709.
\end{equation}
For $E<E_c$ the system takes a core-halo structure and undergoes a
gravitational collapse called Antonov
instability, gravothermal catastrophe, thermal runaway or core
collapse \cite{lbw}. Globular clusters may experience the gravothermal
catastrophe.  The collapse of the core is
self-similar
and leads to a finite-time 
singularity: the central density and the central temperature become infinite in
a finite time while the core radius and the core mass vanish \cite{lbe,cohn}.
The evolution continues in a self-similar postcollapse regime
with the formation of a binary star \cite{inagakilb}. Therefore, the
gravothermal catastrophe  leads
ultimately to the formation of a binary star surrounded by a hot halo. Such a
structure has an infinite entropy $S\rightarrow +\infty$ at
fixed energy (see Appendix A of \cite{sc}).

The caloric curve exhibits a minimum temperature (in the canonical ensemble)
at 
\begin{equation}
\label{hc}
\eta_{c}=2.52, \qquad a'_c=8.99,\qquad {\cal R}'_c=32.1.
\end{equation}
For $T<T_c$ the system undergoes a gravitational collapse called isothermal
collapse \cite{aaiso}. Isothermal stars or self-gravitating Brownian
particles may experience an isothermal collapse. The collapse of the
system is
self-similar
and leads to a finite-time 
singularity: the central density becomes infinite in
a finite time while the core radius and the core mass vanish \cite{sc}.
The evolution continues in a self-similar postcollapse regime
with the formation of a Dirac peak \cite{post}.  Therefore, the isothermal
collapse  leads
ultimately to the formation of a Dirac peak containing all the
mass. Such a structure has an infinite free energy $F\rightarrow
-\infty$ (see Appendix B of \cite{sc}).

In the canonical ensemble the series of equilibria is stable (actually
metastable) on the main
branch until
$\eta_{c}$. According to the Poincar\'e criterion \cite{poincare}, it becomes
unstable
at the first 
turning point of temperature when the specific heat becomes infinite before
becoming negative. A new  mode of instability is lost at each subsequent turning
point of
temperature as the spiral rotates clockwise. In the microcanonical ensemble the
series of equilibria is
stable  (actually metastable) on the main branch until
$\Lambda_{c}$. According to the Poincar\'e criterion \cite{poincare}, it becomes
unstable at the first
turning point of energy when the specific heat vanishes before becoming
positive again. A new  mode of instability is lost at each subsequent turning
point of
energy as the spiral rotates clockwise. There is a region of ensembles
inequivalence, between the turning points of
temperature and energy, i.e., in the first region of negative specific
heat.\footnote{The connection between the sign of the specific heat and the
instability in the canonical  and microcanonical ensembles is explained
in Appendix B of \cite{acb}.}

Dynamical and thermodynamical
stability do not coincide in Newtonian gravity. Thermodynamical
stability implies dynamical stability with respect to the
Vlasov-Poisson equations \cite{ih,aaantonov,cc} but the converse is
wrong. Indeed, it has
been shown that all isotropic stellar systems with a distribution function of
the form $f=f(\epsilon)$ with
$f'(\epsilon)<0$, like the Maxwell-Boltzmann (isothermal)
distribution, are dynamically stable with respect to the Vlasov-Poisson
equations \cite{doremus71,doremus73,gillon76,sflp,ks,kandrup91}, even
those that are thermodynamically unstable.
As a result, the gravothermal catastrophe is a slow (secular) thermodynamical 
instability, not a fast dynamical instability (see Paper I for a more detailed
discussion between dynamical and thermodynamical stability).

\section{Ultrarelativistic limit: the hot spiral}
\label{sec_url}

The ultrarelativistic limit of the classical
isothermal gas corresponds to $k_B T\gg
mc^2$. It can be formally obtained by taking the
limits $T\rightarrow +\infty$,
$b\rightarrow 0$, or $c\rightarrow 0$ in the equations of Sec. \ref{sec_grf}. 

\subsection{Local variables in the ultrarelativistic limit $k_BT\gg mc^2$}
\label{sec_lu}

In the
ultrarelativistic
limit, using $E\simeq pc$, the distribution
function (\ref{a32}) can be written as 
\begin{equation}
\label{lu9}
f({\bf r},{\bf p})=\frac{g}{h^3}e^{\alpha}e^{-pc/k_B T(r)}. 
\end{equation}
From this distribution function, we find that the
local variables $n(r)$, $\epsilon(r)$,
$\epsilon_{\rm kin}(r)$ and $P(r)$ are given by
\begin{equation}
\label{lu2}
n(r)=\frac{8\pi g}{h^3c^3}e^{\alpha}k_B^3 T(r)^3,
\end{equation}
\begin{equation}
\label{lu3}
\epsilon(r)=\epsilon_{\rm kin}(r)=\frac{24\pi
g}{h^3c^3}e^{\alpha}k_B^4T(r)^4,
\end{equation}
\begin{equation}
\label{lu4}
P(r)=n(r)k_B T(r).
\end{equation}
These relations can also be
recovered from the general expressions (\ref{a45})-(\ref{a48}) by using Eq.
(\ref{lu1}).
Combining the foregoing relations, we find that 
\begin{equation}
\label{lu5}
\epsilon(r)=3n(r)k_B T(r),
\end{equation}
\begin{equation}
\label{lu6}
P(r)=\frac{1}{3}\epsilon(r),
\end{equation}
\begin{equation}
\label{lu7}
P(r)=Kn(r)^{4/3},\qquad K=\left (\frac{h^3c^3}{8\pi ge^{\alpha}}\right )^{1/3},
\end{equation}
\begin{equation}
\label{lu8}
n(r)=\left \lbrack\frac{\epsilon(r)}{3K}\right \rbrack^{3/4}.
\end{equation}
On the other hand, using Eq. (\ref{lu2}), the distribution function (\ref{lu9})
can
be written in terms of the density as
\begin{equation}
\label{lu10}
f({\bf r},{\bf
p})=\frac{n(r)c^3}{8\pi k_B^3 T(r)^{3}}e^{-pc/k_B T(r)},
\end{equation}
where we stress that, as in Eq. (\ref{a48b}), the prefactor is actually a
constant. We note that the equation of state (\ref{lu6}) is independent of
$\alpha$
and coincides with the equation of state of the
black-body radiation.\footnote{As shown in Appendix A of Paper
I, this result is valid  for an arbitrary
distribution function in the ultrarelativistic limit.} Furthermore, the relation
(\ref{lu3}) between the energy
density and the temperature may be interpreted as a sort of Stefan-Boltzmann law
$\epsilon=\sigma T^4$ with a ``constant'' $\sigma$ that
depends on $\alpha$. Similarly, the relations (\ref{lu7})
and (\ref{lu8}) between the pressure,  the particle number
density and the energy density are the same as for the black-body radiation
(see Appendix I of Paper I) except that
the constant $K$ depends on $\alpha$. 
These analogies make possible
to use the results obtained in Ref. \cite{aarelat2} for the self-gravitating
black-body radiation in general relativity. However, because of the presence of
$\alpha$
in certain equations, the caloric curve obtained for a classical isothermal gas
in the
ultrarelativistic limit will be different from the  caloric curve of the
self-gravitating radiation obtained in Fig. 15 of Ref. \cite{aarelat2} as
detailed below.

\subsection{General relativistic Emden equations}
\label{sec_gre}

The equilibrium states of a system described by a linear equation of state
$P=q\epsilon$ in general relativity have been studied
in \cite{aarelat1,aarelat2} following the original paper of
Chandrasekhar \cite{chandra72}. Since the equation of state
(\ref{lu6})
is linear (with $q=1/3$) we can directly apply these results to the present
situation. Let us define the
dimensionless
variables $\xi$, $\psi(\xi)$ and $M(\xi)$ by the relations
\begin{equation}
\epsilon=\epsilon_{0}e^{-\psi}, \qquad r=\left ( \frac{c^{4}}{16\pi
G\epsilon_{0}}\right )^{1/2}\xi,
\label{gre2}
\end{equation}
and
\begin{equation}
M(r)=\frac{4\pi\epsilon_{0}}{c^{2}}\left ( \frac{c^{4}}{16\pi
G\epsilon_{0}}\right )^{3/2}M(\xi),
\label{gre3}
\end{equation}
where $\epsilon_0$ represents the energy density at the centre of the
configuration. 
In terms of these variables, the TOV equations (I-99) and (I-104) can be reduced
to the following
dimensionless forms
\begin{equation}
\frac{dM}{d\xi}=\xi^{2}e^{-\psi},
\label{gre5}
\end{equation}
\begin{eqnarray}
\left\lbrack 1-\frac{M(\xi)}{2\xi}\right\rbrack
\frac{d\psi}{d\xi}=\frac{M(\xi)}{\xi^{2}}+\frac{1}{3}\xi e^{-\psi},
\label{gre4}
\end{eqnarray}
with the boundary conditions $\psi(0)=\psi'(0)=0$. These equations are
sometimes called the general relativistic Emden
equations \cite{aarelat1,aarelat2,chandra72}. If we denote by $a$ the value
of
$\xi$ at the edge of the box,\footnote{This
quantity was denoted $\alpha$ in Ref. \cite{aarelat2}. Here, we use the notation
$a$ instead of $\alpha$ to avoid confusion with the variable $\alpha=\mu/k_B T$
introduced
in Sec. \ref{sec_grf}.} we have
\begin{equation}
R=\left ( \frac{c^{4}}{16\pi
G\epsilon_{0}}\right )^{1/2} a\qquad {\rm and}\qquad \xi=\frac{a}{R}r.
\label{gre6}
\end{equation}

\subsection{Particle number}

According to Eqs. (\ref{lu8}), (\ref{gre2}) and (\ref{gre6}),
the 
particle number (\ref{b127}) is given by
\begin{equation}
\label{gre10}
N=4\pi R^3 \left ( \frac{c^4}{48\pi GKR^2}\right )^{3/4}\Delta(a)
\end{equation}
with
\begin{equation}
\label{gre11}
\Delta(a)=\frac{1}{a^{3/2}}\int_0^{a}e^{-3\psi(\xi)/4}\left\lbrack
1-\frac{M(\xi)}{2\xi}\right\rbrack^{-1/2}\xi^2\, d\xi.
\end{equation}
Using the expression of $K$ from Eq. (\ref{lu7}), we obtain
\begin{equation}
\label{gre12}
N=4\pi \left ( \frac{c^4}{48\pi G}\right )^{3/4}  \left ( \frac{8\pi
g}{h^3c^3}\right )^{1/4} R^{3/2}e^{\alpha/4}   \Delta(a).
\end{equation}
Introducing the compactness parameter (\ref{a59}), we can rewrite Eq.
(\ref{gre12})  as
\begin{equation}
\label{gre14}
\nu=4\pi G m \left ( \frac{1}{48\pi G}\right )^{3/4}  \left ( \frac{8\pi
g c}{h^3}\right )^{1/4} R^{1/2}e^{\alpha/4}   \Delta(a).
\end{equation}
For a prescribed value of $\nu$, this equation gives the relation between
$\alpha$ and
$a$.

\subsection{Mass}

According to Eqs. (\ref{gre3}) and
(\ref{gre6}), the mass is given by
\begin{equation}
\label{gre7}
\frac{2GM}{Rc^2}=\chi(a)
\end{equation}
with
\begin{equation}
\label{gre8}
\chi(a)=\frac{M(a)}{2a}.
\end{equation}
Using Eqs. (\ref{a59}) and (\ref{gre7}),
the normalized energy  (\ref{a63}) can be written as
\begin{equation}
\label{gre16}
\Lambda=-\frac{\frac{1}{2}\chi(a)-\nu}{\nu^2}.
\end{equation}

\subsection{Tolman temperature}

The Tolman temperature is given by Eq. (\ref{b130}).
Using Eqs. (\ref{lu3}), (\ref{gre2}), (\ref{gre6}) and (\ref{gre7}), it can be
written as
\begin{equation}
\label{gre18}
k_B T_{\infty}=\left ( \frac{h^3c^7}{384\pi^2 g G R^2}\right
)^{1/4}e^{-\alpha/4}\theta(a)
\end{equation}
with
\begin{equation}
\label{gre19}
\theta(a)=\frac{a^{1/2}}{{\cal R}(a)^{1/4}}\left\lbrack
1-\chi(a)\right\rbrack^{1/2},
\end{equation}
where
\begin{equation}
\label{gre20}
{\cal R}(a)= \frac{\epsilon_0}{\epsilon(R)}=e^{\psi(a)}
\end{equation}
is the energy density contrast. Eliminating $\alpha$ from Eq. (\ref{gre18}) with
the aid of Eq.
(\ref{gre14}), we find that the normalized inverse Tolman temperature
(\ref{a62}) 
is given by
\begin{equation}
\label{gre22}
\eta=\frac{12\nu^2}{\theta(a)\Delta(a)}.
\end{equation}

\subsection{Entropy and free energy}

The entropy is given by Eq. (\ref{sgr}). In the ultrarelativistic limit, using
Eqs. (\ref{a31}) and (\ref{lu5}), we find that 
\begin{equation}
\label{gre22b}
\frac{S}{Nk_B}=4-\alpha.
\end{equation}
The free energy is given by Eq. (\ref{a21WQ}). In the ultrarelativistic limit,
using Eqs. (\ref{a62}), (\ref{a63}) and (\ref{gre22b}), we find that 
\begin{equation}
\label{gre22c}
\frac{FR}{GN^2m^2}=-\Lambda-(4-\alpha)\frac{1}{\eta}.
\end{equation}

\subsection{The caloric curve $\eta(\Lambda)$}

The functions $\chi(a)$, $\Delta(a)$, $\theta(a)$ and ${\cal R}(a)$ can be
obtained by solving the general relativistic Emden equations  (\ref{gre5})
and (\ref{gre4}) numerically. They
have
been plotted in Ref. \cite{aarelat2} and they display damped
oscillations. From these
functions, using Eqs. (\ref{gre16}) and (\ref{gre22}), we can obtain
the caloric curve (series of equilibria)
$\eta(\Lambda)$ of the classical ultrarelativistic gas for
a given value of $\nu$. It has the form of a spiral parametrized by
$a$ or, equivalently, by the energy density contrast ${\cal
R}=\epsilon(0)/\epsilon(R)=e^{\psi(a)}$. It is similar, but not
identical (see below), to the
caloric curve of the self-gravitating black-body radiation plotted in Fig 15 of
\cite{aarelat2}. It
corresponds to
the hot spiral of the general relativistic caloric curve from Fig.
\ref{kcal_N01_linked_colorsPH} when $\nu\rightarrow 0$ and $\alpha\rightarrow
-\infty$ (see Eq. (\ref{gre14}) and Ref.
\cite{acb}). The
first turning point of energy occurs at  
\begin{equation}
\label{gre23}
\Lambda_{\rm min}=-\frac{\frac{1}{2}\chi(a_c)-\nu}{\nu^2},
\end{equation}
where $a_c$ is the value of $a$ corresponding to the first turning point of the
function
$\chi(a)$. It is found in \cite{aarelat2} that $a_c=4.7$ and  $\chi(a_c)=0.493$.
At
that point, the density contrast is ${\cal R}_c=22.4$. Therefore, the value of
$\Lambda_{\rm min}$ for the ultrarelativistic classical self-gravitating gas can
be directly
understood from the results obtained in the context of the self-gravitating
black-body radiation \cite{aarelat2}. On the other hand, the
first turning point of temperature occurs at  
\begin{equation}
\label{gre24}
\eta_{\rm min}=\frac{12\nu^2}{\theta(a'_c)\Delta(a'_c)},
\end{equation}
where $a'_c$ is the value of $a$
corresponding to the first turning point of the
function
$\theta(a)\Delta(a)$. We note that the maximum temperature $(T_{\infty})_{\rm
max}$ of the ultrarelativistic classical self-gravitating gas is different from
the maximum
temperature $(T_{\infty})_{\rm max}$ of the
self-gravitating radiation discussed in Sec. 3.5 of \cite{aarelat2} which
corresponds to the first turning point of the function
$\theta(a)$. We find that $a'_c=3.48$ and
$\theta(a'_c)\Delta(a'_c)=0.674$. At that
point, the density contrast is ${\cal R}'_c=10.3$ (by
comparison, in Ref. \cite{aarelat2}, we found  $a''_c=1.47$, 
$\theta(a''_c)=0.897$ and ${\cal R}''_c=1.91$).

\subsection{The asymptotic caloric curve ${\cal B}({\cal M})$}

\begin{figure}
\begin{center}
\includegraphics[clip,scale=0.3]{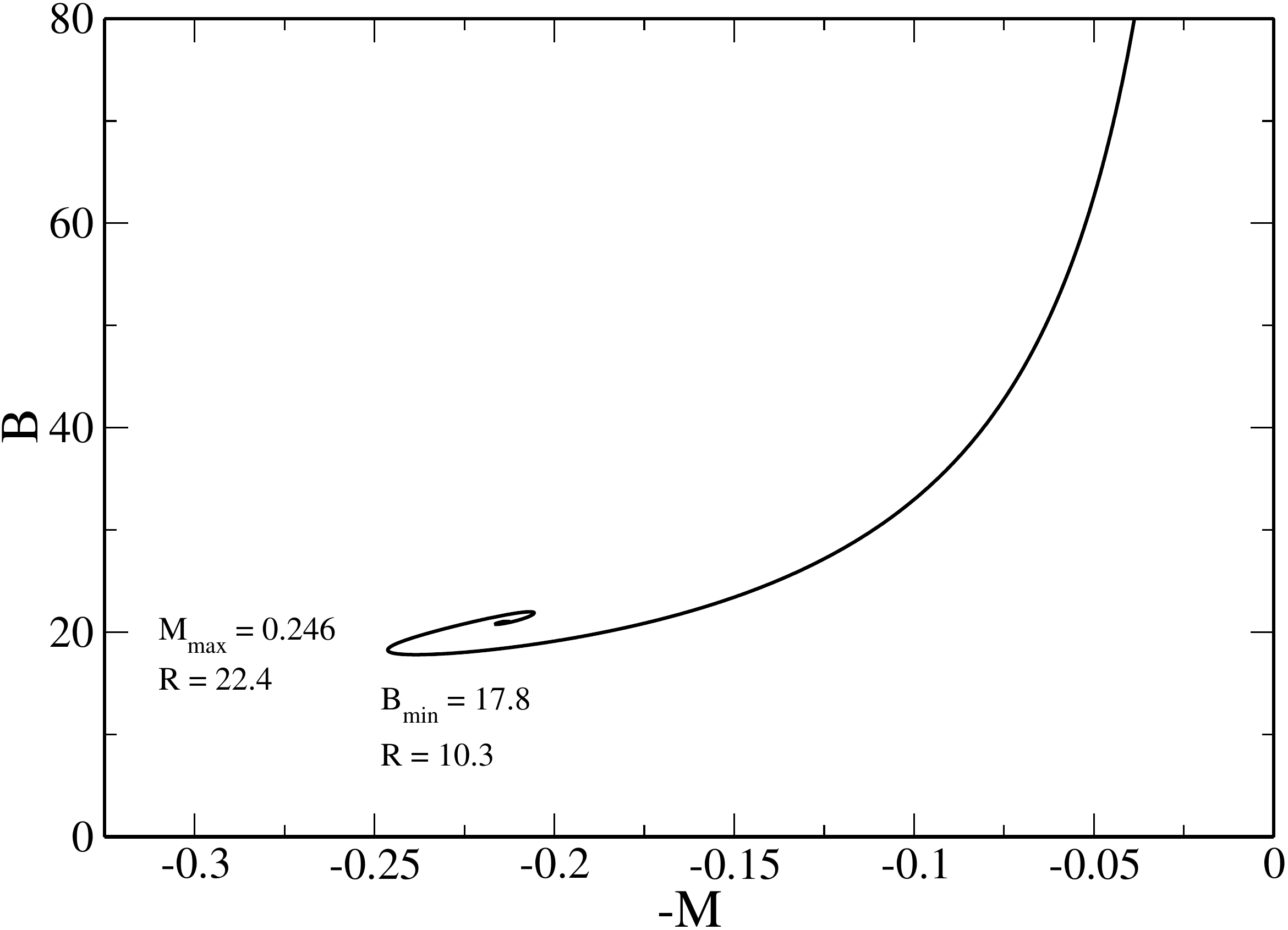}
\caption{Caloric curve ${\cal B}({\cal M})$ of the ultrarelativistic
self-gravitating  classical gas obtained by solving the general relativistic
Emden
equations (\ref{gre5})
and (\ref{gre4}). It corresponds to the limit curve of the general
relativistic caloric curve from Fig.
\ref{kcal_N01_linked_colorsPH} when $\nu\rightarrow 0$ using the variables
${\cal M}$ and ${\cal B}$ instead of $\Lambda$ and $\eta$. }
\label{hot}
\end{center}
\end{figure}

The previous results are valid in the
ultrarelativistic limit of the general
relativistic
classical isothermal gas, corresponding to
$\nu\rightarrow 0$ and $\alpha\rightarrow -\infty$. In
that limit, we see from Eqs. (\ref{gre23}) and (\ref{gre24}) that 
$\Lambda_{\rm min}\rightarrow -\infty$ and $\eta_{\rm min}\rightarrow 0$. More
precisely,
\begin{equation}
\label{gre23b}
\Lambda_{\rm min}\sim -\frac{0.246-\nu}{\nu^2}\qquad {\rm and}\qquad \eta_{\rm
min}\sim 17.8\, \nu^2.
\end{equation}
Therefore, in terms of the variables $\Lambda$ and $\eta$, the hot
spiral is rejected at infinity when $\nu\rightarrow 0$. This is the conclusion
we had reached in Sec.
\ref{sec_cold} when studying the nonrelativistic limit. The  
nonrelativistic limit corresponds to small values of $E$ and
$T$ such that $\Lambda,\eta\sim 1$ when $\nu\rightarrow 0$. By contrast,  the
ultrarelativistic limit
corresponds to large values of $E$ and $T$ such that $\Lambda\rightarrow
-\infty$ and $\eta\rightarrow 0$ when $\nu\rightarrow 0$. We need therefore  to
introduce new scales in
order to scan this part of the caloric curve. According to Eqs.
(\ref{gre16})
and (\ref{gre22}), the caloric
curve $\eta(\Lambda)$ presents a clear
scaling. Indeed, when $\nu\rightarrow
0$, the caloric curve defined in terms of  the
variables $\eta/\nu^2$ and $\Lambda\nu^2-\nu$ (instead of $\eta$ and $\Lambda$)
tends towards a limit curve given in parametric form by
\begin{equation}
\label{gre25}
\Lambda\nu^2-\nu=-\frac{1}{2}\chi(a)\qquad {\rm and}\qquad 
\frac{\eta}{\nu^2}=\frac{12}{\theta(a)\Delta(a)}.
\end{equation} 
This prompts us to introducing the new dimensionless energy and
temperature variables relevant to the ultrarelativistic limit
\begin{equation}
\label{gre26b}
{\cal M}= \frac{GM}{Rc^2}\qquad {\rm and}\qquad 
{\cal B}=\frac{Rc^4}{GN k_B T_{\infty}}.
\end{equation} 
They are related to $\Lambda$ and $\eta$ by
\begin{equation}
\label{gre25w}
{\cal
M}=-(\Lambda\nu^2-\nu)\qquad  {\rm and}\qquad {\cal
B}=\frac{\eta}{\nu^2},
\end{equation} 
or inversely,
\begin{equation}
\label{gre25wi}
\Lambda= -\frac{{\cal M}-\nu}{\nu^2}\qquad {\rm and}\qquad \eta= {\cal B}\,
\nu^2.
\end{equation} 
In Eq. (\ref{gre26b}) the
mass is normalized by $M_*=Rc^2/G$ (similar to the Schwarzschild mass) and the
Tolman temperature is
normalized by $k_B T_*=Rc^4/GN$ (similar to the Schwarzschild energy
per particle). The mass scale $M_*$ is the same as the one introduced for the
self-gravitating radiation \cite{aarelat2}. By contrast, the
temperature scale $T_*$ is different from the
temperature scale of the self-gravitating radiation
$k_B T_*^{\rm rad}=(\hbar^3c^7/GR^2)^{1/4}$ introduced in  Sec. 3.5
of \cite{aarelat2} which
depends on $\hbar$. 
The caloric
curve  ${\cal B}({\cal M})$ for different values of $\nu$ is represented in
Figs. 20 and 21 of \cite{acb}. For $\nu\rightarrow 0$, it tends
towards a limit curve which,
according to Eq. (\ref{gre25}), is given by 
\begin{equation}
\label{gre26}
{\cal M}=\frac{1}{2}\chi(a)\qquad {\rm and}\qquad 
{\cal B}=\frac{12}{\theta(a)\Delta(a)}.
\end{equation} 
This limit curve is represented in
Fig.
\ref{hot}.  It has the form of a spiral
along which the energy density contrast ${\cal R}$ increases.  It
corresponds to
the hot spiral of the general relativistic caloric curve from Fig.
\ref{kcal_N01_linked_colorsPH} when $\nu\rightarrow 0$ and $\alpha\rightarrow
-\infty$ (see Eq. (\ref{gre14}) and Ref.
\cite{acb}).  In that limit, in which $k_B T \gg mc^2$, the cold spiral is
rejected at
infinity (${\cal M}_c\rightarrow 0$ and ${\cal B}_c\rightarrow +\infty$) when
we use the variables ${\cal M}$ and ${\cal B}$. More
precisely, using Eqs. (\ref{lc}) and (\ref{hc}), we get
\begin{equation}
\label{gre23bb}
{\cal M}_{c}\sim \nu-0.335\, \nu^2 \qquad {\rm and}\qquad  {\cal B}_{c}\sim
\frac{2.52}{\nu^2}.
\end{equation}
The caloric curve ${\cal B}({\cal M})$ exhibits a maximum energy
(in the microcanonical ensemble) and a
maximum temperature (in the canonical ensemble)  at 
\begin{equation}
\label{gre27}
{\cal M}_{\rm max}=\frac{GM_{\rm
max}}{Rc^2}=\frac{1}{2}\chi(a_c)=0.24632,\qquad
a_c=4.70, \qquad 
{\cal R}_c=22.4,
\end{equation}
\begin{equation}
\label{gre28}
{\cal B}_{\rm min}=\frac{Rc^4}{GNk_B (T_{\infty})_{\rm
max}}=\frac{12}{\theta(a'_c)\Delta(a'_c)}=17.809\qquad
a'_c=3.48 \qquad 
{\cal R}'_c=10.3.
\end{equation} 
For $M>M_{\rm max}$ (in the microcanonical ensemble) or
$T_{\infty}>(T_{\infty})_{\rm max}$ (in the canonical ensemble) the
system undergoes a gravitational collapse leading
presumably to the formation of a black hole.

In the canonical ensemble the series of equilibria is stable (actually
metastable) on the main
branch until
${\cal B}_{\rm min}$. According to the Poincar\'e criterion \cite{poincare}, it
becomes
unstable
at the first 
turning point of temperature when the specific heat becomes infinite before
becoming negative. A new  mode of instability is lost at each subsequent turning
point of
temperature as the spiral rotates clockwise. In the microcanonical ensemble the
series of equilibria is
stable  (actually metastable) on the main branch until
${\cal M}_{\rm max}$. According to the Poincar\'e criterion \cite{poincare}, it
becomes
unstable at the first
turning point of mass-energy when the specific heat vanishes before becoming
positive again. A new  mode of instability is lost at each subsequent turning
point of
mass-energy as the spiral rotates clockwise. There is a region of ensembles
inequivalence between the turning points of
temperature and mass-energy, i.e., in the first region of negative specific
heat.

The caloric curve ${\cal B}({\cal M})$ is similar, but not identical, to the
caloric curve of the self-gravitating black-body radiation represented
in Fig. 15 of \cite{aarelat2}. The maximum mass $M_{\rm max}$ is the same
(given by Eq. (\ref{gre27})) but the maximum temperature $(T_{\infty})_{\rm
max}$ is fundamentally different. In the case of the self-gravitating radiation,
one has ${\cal T}_{\rm max}={k_B (T_{\infty})_{\rm max}
G^{1/4}R^{1/2}}/{\hbar^{3/4}c^{7/4}}=0.445$ and ${\cal R}''_c=1.91$ instead of
Eq. (\ref{gre28}).

According to Ipser's conjecture
\cite{ipser80} (see Paper I and footnote 5), dynamical and thermodynamical
stability coincide in general relativity. Therefore, the collapse at $M_{\rm
max}$ is essentially a dynamical instability which takes place on a short
timescale.

{\it Remark:} According to Eq. (\ref{lu3}), we have the relation
$\epsilon(r)\propto T(r)^4$ between the energy density and the local temperature
for a given equilibrium state (specified by $\alpha$). Therefore, if we define
the temperature contrast by $\Theta=T(0)/T(R)$ \cite{acb} we find that
\begin{equation}
\label{rtlast}
{\cal R}=\frac{\epsilon(0)}{\epsilon(R)}=\frac{T(0)^4}{T(R)^4}=\Theta^4.
\end{equation}
This yields $\Theta_c={\cal R}_c^{1/4}=2.18$ and $\Theta'_c={{\cal
R}'_c}^{1/4}=1.79$. The relation (\ref{rtlast}) also holds for the
self-gravitating black-body radiation.

\subsection{The caloric curve $b_{\infty}(e)$}

There are different manners to plot the caloric curve of the general
relativistic classical gas. We have previously discussed the representations
$\eta(\Lambda)$ and ${\cal B}({\cal M})$. We could have also introduced the
normalizations
\begin{equation}
\label{nn1}
\frac{1}{b_{\infty}}=\frac{k_B T_{\infty}}{mc^2}=\frac{\nu}{\eta}=\frac{1}{{\cal
B}\nu} \qquad {\rm and} \qquad
e=\frac{E}{Nmc^2}=\frac{M}{Nm}-1=-\Lambda\nu=\frac{{\cal M}}{\nu}-1
\end{equation}
for the temperature and the energy ($e$ is the
fractional binding energy). We note that these normalized variables do not
depend on the
box radius $R$. For $\nu\rightarrow 0$, the turning points of the
cold spiral behave as
\begin{equation}
\label{nn3}
\frac{1}{b^c_{\infty}}=\frac{k_B T_{\infty}^c}{mc^2}\sim \frac{\nu}{2.52},
\end{equation}
\begin{equation}
\label{nn4}
e_c=\frac{E_c}{Nmc^2}=\frac{M_c}{Nm}-1\sim -0.335 \nu,
\end{equation}
while the turning
points of the hot spiral behave as
\begin{equation}
\label{nn5}
\frac{1}{b^{\rm min}_{\infty}}=\frac{k_B T_{\infty}^{\rm max}}{mc^2}\sim
\frac{1}{17.809\nu},
\end{equation}
\begin{equation}
\label{nn6}
e_{\rm max}=\frac{E_{\rm max}}{Nmc^2}=\frac{M_{\rm max}}{Nm}-1\sim
\frac{0.24632}{\nu}-1.
\end{equation}
We note
that the caloric curve giving $b_{\infty}=mc^2/k_B T_{\infty}$ as a
function of
$e={E}/{Nmc^2}={M}/{Nm}-1$ does
not tend to a limit when $\nu\rightarrow 0$. Therefore, the representations 
$\eta(\Lambda)$ and ${\cal B}({\cal M})$ seem to be more adapted to our problem
than the 
representation $b_{\infty}(e)$.

\section{Conclusion}

In this paper, using  the formalism of Paper I, we have studied the statistical
mechanics of classical self-gravitating systems within the framework of general
relativity. The equations derived in this paper allow us to understand the 
construction of the caloric curves of classical self-gravitating systems
obtained in Newtonian
gravity
\cite{antonov,lbw,katzpoincare1,lecarkatz,paddyapj,paddy,katzokamoto,
dvs1,dvs2,aaiso,crs,sc,grand,katzrevue,lifetime,ijmpb} and
general relativity \cite{roupas,acb}. Generically, the caloric curve has the
form of a double spiral. The turning points of temperature and energy on the
spirals reflect the occurrence of a gravitational collapse in the
canonical and microcanonical ensembles
respectively. At low
temperatures, the gas collapses because
it is ``too cold'' to provide the thermal pressure necessary to equilibrate
self-gravity. At high energies the gas collapses because it is ``too hot'' and
feels  the ``weight of heat'' \cite{tolman}. We have investigated precisely the
nonrelativistic and ultrarelativistic limits of the classical self-gravitating
gas.

The nonrelativistic limit corresponds to $\nu=GNm/Rc^2\rightarrow 0$ with
$\eta=GNm^2/k_B TR\sim 1$ and $\Lambda=-ER/GN^2m^2\sim 1$. The normalized
variables $\eta$ and $\Lambda$ are adapted to scan ``low''
values of temperature and energy. In the nonrelativistic limit, using the
representation $\eta(\Lambda)$, the hot spiral
is rejected at 
infinity and only the cold spiral remains.  We
have $\Lambda_c\rightarrow 0.335$
and $\eta_c\rightarrow 2.52$ while $\Lambda_{\rm min}\sim -(0.246-\nu)/\nu^2$
and $\eta_{\rm min}\sim 17.8\, \nu^2$. We also have $\alpha\rightarrow
+\infty$   and $k_B T\ll mc^2$. Therefore, we obtain the limit
curve of Fig. \ref{etalambda} displaying the nonrelativistic cold spiral. This
asymptotic caloric curve first appeared in  \cite{katzpoincare1}. The
convergence towards that curve when $\nu\rightarrow 0$ is shown in Figs. 15 and
16 of \cite{acb}.

The ultrarelativistic limit corresponds to $\nu=GNm/Rc^2\rightarrow
0$ with
${\cal M}=GM/Rc^2\sim 1$ and ${\cal B}=Rc^4/GNk_B T_{\infty}\sim 1$. The
normalized variables ${\cal M}$ and ${\cal B}$ are adapted to scan ``large''
values of
temperature and energy. In that limit, , using the
representation ${\cal B}({\cal M})$, the cold spiral is rejected at 
infinity and only the hot spiral remains. We have ${\cal M}_{\rm
max}\rightarrow 0.24632$ and ${\cal B}_{\rm min}\rightarrow 17.809$ while 
${\cal
M}_{c}\sim \nu-0.335\nu^2$ and ${\cal B}_{c}\sim 2.52/\nu^2$.  We also
have $\alpha\rightarrow -\infty$ and $k_B T\gg mc^2$. Therefore, we obtain the
limit
curve of Fig.  \ref{hot} displaying the ultrarelativistic hot spiral.
This asymptotic
caloric curve is new. The convergence
towards that curve when $\nu\rightarrow 0$ is shown in Figs. 20 and 21 of
\cite{acb}. We have also discussed the analogies and the differences between
this asymptotic caloric curve and the caloric curve of the self-gravitating
black body radiation obtained in Fig. 15 of \cite{aarelat2}.

\appendix

\section{Series of equilibria of truncated isothermal distributions}
\label{sec_scc}

In this Appendix, we discuss the series of equilibria of truncated isothermal
distributions in Newtonian gravity and general relativity.

\subsection{Nonrelativistic systems}
\label{sec_sccnr}

\begin{figure}
\begin{center}
\includegraphics[clip,scale=0.3]{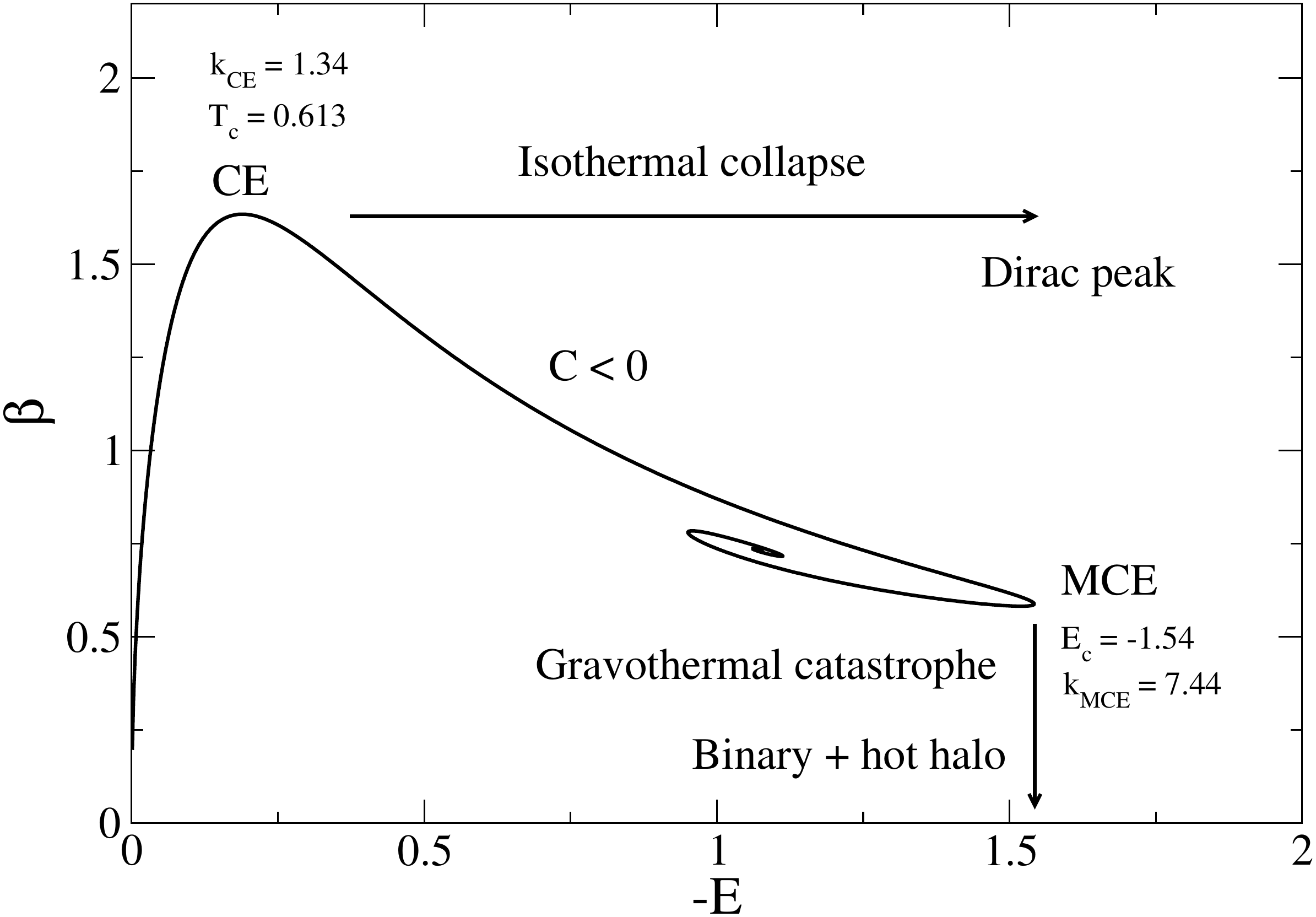}
\caption{Series of equilibria (caloric curve) of the classical King model. 
The units for the curve $\beta(E)$ are $(m v_0^{2})^{-1}$ for $\beta$ and
$Mv_0^2$ for
$E$ where $v_0$ is a typical velocity defined in \cite{katzking,clm1}. The
series of equilibria is parametrized by the dimensionless concentration
parameter $k=\beta[\Phi(R)-\Phi(0)]$. It has a snail-like
(spiral) structure but only the part of the curve up to CE ($k_{\rm CE}=1.34$)
is stable in the canonical ensemble and only the part up to  MCE ($k_{\rm
MCE}=7.44$) is stable in the
microcanonical ensemble (the region between CE and MCE where the specific
heat is negative
corresponds to a region of ensembles inequivalence). For
$k\rightarrow 0$ the King model is equivalent to a polytrope of index $n=5/2$
\cite{katzking,clm1} that is both
canonically and microcanonically stable \cite{aaantonov} (the same is true for
the Wooley model which is equivalent, for $k\rightarrow 0$, to a
polytrope $n=3/2$ \cite{katzking}). The
equilibrium
states are all dynamically (Vlasov) stable.
}
\label{ebAintro}
\end{center}
\end{figure}

The series of equilibria  of globular clusters described by
truncated isothermal distributions (Woolley \cite{woolley}
and King \cite{king} models) have been
determined by Lynden-Bell and Wood
\cite{lbw}, Katz \cite{katzking} and
Chavanis {\it et al.} \cite{clm1}. The caloric curve $\beta(E)$ of the King
model is reproduced in Fig.
\ref{ebAintro}. It has the form of a spiral. It is parametrized by the
concentration parameter $k$ that increases monotonically along the series of
equilibria. The curves $\beta(k)$ and $E(k)$ giving
the inverse temperature and the energy as a function of the concentration
parameter $k$
display damped oscillations \cite{clm1}.

The thermodynamical stability of
truncated isothermal distributions was analyzed in Refs.
\cite{lbw,katzking,clm1} by using the Poincar\'e turning point criterion
\cite{poincare}. For $E\rightarrow 0^-$
and $\beta\rightarrow 0$, we know that the system is stable because it
is equivalent to a polytrope of index $n=5/2$ that is both canonically and
microcanonically stable \cite{aaantonov}. In the
canonical ensemble (fixed temperature), the series of equilibria is stable up to
the first turning point of temperature (corresponding to $k_{\rm CE}=1.34$) and
becomes
unstable afterwards. This is when the specific heat becomes infinite,
passing from positive to negative values. In the microcanonical ensemble (fixed
energy) the series
of
equilibria is stable up to the first turning point of energy (corresponding
to $k_{\rm
MCE}=7.44$) and
becomes unstable afterwards. This is when the specific heat vanishes,
passing from negative to positive values. The
statistical ensembles are inequivalent in the region of negative specific
heat ($C<0$), between the turning point of temperature CE and the turning point
of energy MCE.
From general arguments, it can be shown that canonical
stability implies microcanonical stability \cite{aaantonov,cc}. Basically,
this
is because the microcanonical ensemble is more constrained, hence more stable,
than the canonical ensemble. In the present case, this manifests itself (in
conjunction with the Poincar\'e theory)  by the fact that the turning point
of temperature occurs before the turning point of energy. For isolated stellar
systems, only the
microcanonical
ensemble makes sense physically.\footnote{We note that globular clusters become
rapidly
canonically unstable, i.e., as soon as they are substantially different from
a polytrope $n=5/2$
(see the discussion in \cite{clm1}). This is a clear sign of the fact that real
globular clusters are described by the microcanonical ensemble in which they are
stable longer. Indeed, if they were described by the canonical ensemble, most
of the observed globular clusters would be unstable.}

Let us now consider the dynamical stability of the system with respect to a
collisionless evolution described by the Vlasov-Poisson equations. In Newtonian
gravity, it has been shown that all isotropic stellar systems with a
distribution function
of the form $f=f(\epsilon)$ with $f'(\epsilon)<0$ are
dynamically stable \cite{doremus71,doremus73,gillon76,sflp,ks,kandrup91}.
Therefore, the whole series of
equilibria of isothermal stellar systems is dynamically
stable, even the equilibrium states deep into the spiral that are
thermodynamically unstable. We know from general
arguments that thermodynamical stability
implies dynamical stability \cite{ih,cc}. 
The results of \cite{doremus71,doremus73,gillon76,sflp,ks,kandrup91}
show that the converse is wrong in Newtonian
gravity:
before the first turning point of energy, the system is both 
thermodynamically stable (in the microcanonical ensemble) and dynamically
stable; after the first turning point of energy, the system is 
thermodynamically unstable while it is still dynamically stable.

We also know from general
arguments that the thermodynamical stability of a stellar system in the
canonical ensemble  is equivalent to
the dynamical stability of the corresponding barotropic star with respect to
the Euler-Poisson equations \cite{aaantonov}.
Therefore, barotropic stars with the equation of state
corresponding to the truncated isothermal distribution function are
dynamically stable before the first turning point of temperature and 
dynamically
unstable after the first turning point of temperature. 

The dynamical
evolution of globular clusters is discussed in
the introduction (see also Appendix D of \cite{clm1}). Because of collisions
between stars and evaporation, the system evolves quasi-statically along the
series
of equilibria. The natural evolution corresponds to an increase of the central
density $\rho_0$ that parametrizes the series
of equilibra.\footnote{This can be understood from different
arguments: (i) Under the
effect of close encounters,
stars leave the system with an energy positive or close to zero. Therefore, the
energy of the cluster decreases or remains approximately constant. Since the
number of stars in the cluster decreases, the cluster contracts (according to
the virial theorem) and becomes more and more concentrated. 
(ii) For globular
clusters described by the King model, one can show that the Boltzmann entropy
$S_B$ is an increasing function of the concentration parameter $k$ until a point
$k_{*}$ at which the Boltzmann entropy reaches a maximum before decreasing (see
Table II of \cite{lbw}, Fig. 5 of \cite{cohn} and Fig. 46 of \cite{clm2}).
Therefore, we can relate the temporal increase of the concentration parameter
$k(t)$ on the series of equilibria with the second principle of thermodynamics,
i.e., the temporal increase of the Boltzmann entropy ${\dot S}_B\ge 0$ 
($H$-theorem). This adiabatic evolution continues (at most) until the point
$k_{*}$
at which the Boltzmann entropy is maximum since the Boltzmam entropy cannot
decrease with time. At the instability point $k_{\rm
MCE}$, the system becomes unstable, undergoes the gravothermal catastrophe, and
evolves away from the series of equilibria. The instability point
$k_{\rm MCE}$, corresponding to the extremum of the King entropy $S$
(defined in \cite{clm2}) or equivalently  to the 
turning point of energy (since $\delta S=\beta \delta E$),  occurs a bit sooner
than the point $k_*$ at which
the Boltzmann entropy is maximum (see the discussion in \cite{clm2}). It is a
bit disturbing to note that the King entropy
decreases as the concentration parameter increases along the series of
equilibria (see Fig. 46 of \cite{clm2}). There is, however, no paradox since the
$H$-theorem applies to the Boltzmann entropy not to the King entropy. On the
other hand, for
box-confined systems,
the Boltzmann entropy is a decreasing function of the density contrast (see Fig
3 of
\cite{pt}) but, in that case, the system does not evolve along the series of
equilibria so there is no paradox either.} During the evolution, the energy
decreases. On the other hand, the
temperature decreases in the region of positive specific heat $C=dE/dT>0$ and
increases in the region of negative specific heat $C=dE/dT<0$. At the first
turning point of energy
(minimum energy state) the system becomes unstable and undergoes the
gravothermal catastrophe (core collapse). This is a thermodynamical instability
taking place on a relaxation (secular) timescale. It ultimately leads to the
formation of a binary star surrounded by a hot halo.

\subsection{General relativistic systems}
\label{sec_sccr}

\begin{figure}
\begin{center}
\includegraphics[clip,scale=0.3]{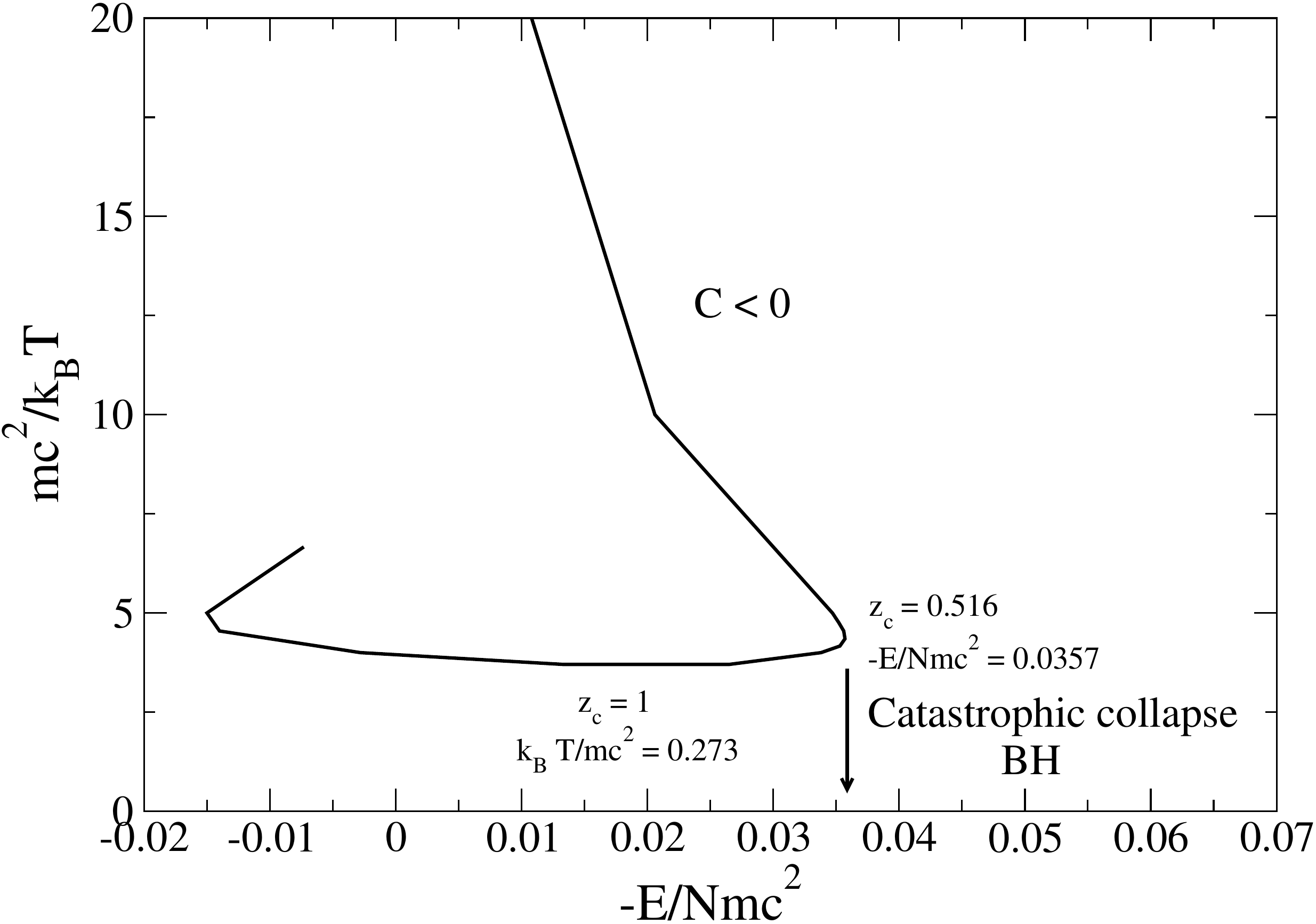}
\caption{Series of equilibria (caloric curve) of heavily
truncated Maxwell-Boltzmann
distributions in general relativity (according to the data of
Ipser \cite{ipser69b}).
The
series of equilibria is parametrized by the central redshift $z_0$. It has
a snail-like
(spiral) structure of which we only see the begining. The equilibrium states are
all canonically unstable. In the microcanonical ensemble, only
the
part of the curve up to $z_c=0.516$ (corresponding to the turning point of
binding energy) is
stable. In this region the specific heat is negative.  
The onset of thermodynamical instability coincides with
the onset of dynamical instability.}
\label{ipser}
\end{center}
\end{figure}

The series of equilibria of relativistic star clusters
described by the heavily
truncated Maxwell-Boltzmann distribution (relativistic Woolley model) was
first determined  by Zel'dovich and
Podurets \cite{zp}. They plotted the temperature measured by an
observer at infinity $T_{\infty}$ as a function of the central density
$\rho_0$ and found that the curve $T_{\infty}(\rho_0)$ dispays damped
oscillations.\footnote{They mentioned that the damped oscillations of
$T(\rho_0)$ are similar to the damped oscillations of $M(\rho_0)$ for neutron
stars discovered by Dmitriev and Kholin \cite{dk}.
This is because, in the
ultrarelativistic limit where the density (or the redshift) is large, the
equation of state of a
classical
isothermal gas takes the form $P=\epsilon/3=K n^{4/3}$, where $\epsilon$ is the
energy
density and $n$ the particle number, like the ultrarelativistic equation of
state of a Fermi gas at $T=0$ (or like the black-body radiation).}
As a result, there exists a
maximum temperature $k_B T_{\rm max}/mc^2=0.273$ above which there is no
equilibrium.\footnote{Zel'dovich and Podurets
\cite{zp} assumed a certain relation between the 
energy cutoff and the 
temperature. This relatively {\it ad hoc} choice was later criticized. This led
to several
generalizations of the problem by Katz {\it et al.} \cite{khk}, Suffern
and Fackerell \cite{sf76}, Fackerell and Suffern \cite{fs76}, Merafina
and
Ruffini \cite{mr,mreuro,mr90}, and Bisnovatyi-Kogan {\it et
al.}
\cite{bmrv,bmrv2} that we
do not review in detail here.}  They argued (without rigorous
justification) that the series of equilibria should become unstable at that
point and that the system should experience a  gravitational collapse that they
called an ``avalanche-type
catastrophic contraction of the system''. Considering the same distribution
function, Ipser
\cite{ipser69b}
plotted the fractional
binding energy $E/Nmc^2$ as a function of the central redshift $z_0$  and
found that the curve $E/Nmc^2(z_0)$ displays damped oscillations. Using a
rigorous instability criterion based on the equation of pulsations derived by
Ipser and Thorne \cite{ipserthorne}, he found that the series of
equilibria becomes dynamically (Vlasov) unstable above a critical redshift
$z_c=
0.516$  and that
this critical
value happens to coincide with the turning point of fractional binding energy.
At that point $[(M-Nm)/Nm]_{\rm c}=0.0357$, $(Rc^2/2GNm)_{\rm c}=4.42$ and
$(k_B T_\infty/mc^2)_{\rm
c}=0.23$. The turning
point of binding energy found by Ipser
\cite{ipser69b} is different
from the turning
point of temperature found by  Zel'dovich and
Podurets \cite{zp} corresponding to $z_0=1.08$, $(M-Nm)/Nm=0.0133$,
$Rc^2/2GNm=3.92$, and
$k_B(T_{\infty})_{\rm max}/mc^2=0.27$. In particular, the gravitational
instability
occurs sooner
than
predicted by  Zel'dovich and
Podurets \cite{zp}. Using the values of the fractional binding energy and
of the temperature measured by an observer at infinity tabulated by
Ipser \cite{ipser69b} (see his Table 1), we have
plotted in Fig. \ref{ipser} the caloric curve of general relativistic
heavily truncated Maxwell-Boltzmann distributions giving $mc^2/k_B T_{\infty}$
as a function of $E/Nmc^2$. This curve has not been plotted before. It has the
form of a spiral
of which we only see the beginning.

We can now investigate the dynamical and thermodynamical stability of the
equilibrium states of relativistic star clusters in more detail. Let us first
consider their thermodynamical stability. In a sense, we can say that 
Zel'dovich and Podurets \cite{zp} considered the
canonical ensemble (fixed $T_{\infty}$) while Ipser \cite{ipser69b} considered
the
microcanonical ensemble
(fixed $E$). The fact that they obtained different results is related to the
notion of ensembles inequivalence for systems with long-range interactions like
self-gravitating systems. 

Let us first consider the microcanonical ensemble which is the rigorous
statistical ensemble that we should use in order to describe isolated star
clusters.
Let us assume that the system is
stable at sufficiently high energies (i.e. close to $0^{-}$). Using
the Poincar\'e criterion \cite{poincare}, we conclude that the
system is microcanonically stable on the upper branch of Fig. \ref{ipser} up
to the first
turning point of fractional binding energy and becomes unstable afterwards. We
note that the system becomes
microcanonically unstable when the specific heat becomes positive. This
result was first stated by Horwitz and Katz \cite{hk}.

Let us now turn to the canonical ensemble. If the system were
stable at sufficiently low temperatures, using
the Poincar\'e criterion \cite{poincare}, we would conclude that the
system is canonically stable up to the first
turning point of temperature and becomes canonically unstable afterwards.
However, considering the topology of the series of equilibria in Fig.
\ref{ipser}, this is not
possible. Indeed, we
know from general arguments that canonical stability implies microcanonical
stability 
\cite{aaantonov,cc}. Now, we note that the first turning
point
of
temperature occurs {\it after}
the first turning point of binding energy. 
Therefore, if the series of equilibria were
canonically stable up to the first turning point of temperature it would also be
microcanonically stable up to that point. But, we know that the series of
equilibria becomes unstable after the first turning point of binding energy.
Therefore, we arrive at a contradiction. Since a change of stability in the
canonical ensemble can occur only at a turning point of temperature, we conclude
that the whole series of equilibria is canonically unstable.\footnote{The same
is true in Newtonian gravity for the isotropic Wilson model which presents a
similar series of equilibria (see Fig. 1 of \cite{katzking}). This is confirmed
by the fact that for $k\rightarrow 0$ the Wilson model is equivalent to a
polytrope
of index $n=7/2>3$ that is canonically unstable (the corresponding gaseous
spheres
are dynamically unstable with respect to the Euler-Poisson equation) while
being microcanonically stable (see Ref. \cite{aaantonov} for more details).}

Let us finally consider the dynamical stability of the system with respect to a
collisionless evolution described by the Vlasov-Einstein equations. We know from
general
arguments that thermodynamical stability
implies dynamical stability \cite{ipser80}. From the previous
results obtained in the
microcanonical ensemble, we conclude that the system is both
thermodynamically and dynamically stable up to
the first
turning point of fractional binding energy. This is a particular case (for
isothermal systems) of
the general binding energy criterion derived by Ipser \cite{ipser80}. After the
first
turning point of fractional binding energy the system is thermodynamically
unstable. On the other hand, Ipser \cite{ipser69b} has shown numerically that
the
system is also
dynamically unstable after that point. This lead Ipser \cite{ipser80} to the
conclusion\footnote{Actually, this remains a conjecture because it has not been
proven mathematically that isothermal spheres become unstable
after the turning point of fractional binding energy. This is only a numerical
result valid for heavily truncated isothermal distributions with a certain
relation between the energy cutoff and the temperature. Furthermore, the main
open question is
to know whether this result is true for all isotropic distribution functions,
i.e., if a collisionless relativistic star cluster always becomes dynamically
unstable after the turning point of binding energy. This property has been
observed numerically for all the distributions functions that
have been considered \cite{ipser69b,st2} but there
is no rigorous
proof of this result in general.} that, in
general relativity, dynamical and microcanonical thermodynamical stability
coincide contrary to the case of Newtonian systems (see Appendix
\ref{sec_sccnr}).

From the  previous results obtained in the canonical ensemble, we cannot
conclude
anything regarding the dynamical stability of collisionless isothermal
star clusters since all equilibria are canonically unstable. However, we know
that the canonical stability of a collisionless star cluster is equivalent to
the dynamical stability of the corresponding barotropic star with respect to
the Euler-Einstein equations \cite{roupas1,roupas1E,gsw,fhj}.
Therefore, our
observation that all the equilibria of isothermal relativistic star clusters
are
canonically unstable is consistent with Ipser's \cite{ipser69b} finding
that all the relativistic stars with the same equation of state as the
isothermal star clusters are dynamically
unstable.

The dynamical evolution of relativistic star clusters is
reviewed in the Introduction (see also the introduction of
\cite{acb}). Because of collisions between stars and evaporation, the system
evolves quasi-statically along the series
of equilibria. The natural evolution corresponds to an increase of the central
density $\rho_0$ (or central redshift $z_0$)
that parametrizes the series
of equilibra. During the evolution, the energy decreases. On the other hand, the
temperature decreases in the region of positive specific heat
$C=dE/dT_{\infty}>0$ and
increases in the region of negative specific heat $C=dE/dT_{\infty}<0$. At the
turning point of energy
(minimum energy state) the system becomes unstable and undergoes a
gravitational collapse. This is a dynamical (and thermodynamical) 
instability of general relativistic origin taking place on a dynamical 
(short) timescale. It ultimately leads to the
formation of a black hole surrounded by a halo of stars.

\section{Analogies and differences between box models and truncated
distributions}
\label{sec_sccnrb}

\subsection{Newtonian gravity}
\label{sec_sccnrng}

In Newtonian gravity, the series of equilibria of truncated isothermal
distributions
(see Fig. \ref{ebAintro}) is qualitatively similar to the series of equilibria
of
box-confined isothermal systems (see Fig. \ref{etalambda} and the ``cold
spiral'' of Fig. \ref{kcal_N01_linked_colorsPH}). The main difference
stems from the
fact that, for open clusters, equilibrium states necessarily have a negative
energy $E<0$.\footnote{This property
immediately results from the equilibrium scalar virial
theorem  $2K+W=0$ for an unbounded self-gravitating system implying 
$E=K+W=(1/2)W=-K<0$.} Indeed, when the energy is
positive ($E>0$) the
stars
are unbounded and disperse away. If the system is
confined within
a box,
equilibrium states with a positive energy are possible because the stars
bounce off the wall. Apart
from this difference, the  series of equilibria in Figs. \ref{etalambda} and 
\ref{ebAintro} have a similar
spiralling shape. They display a  turning point of
temperature before the turning point of energy. For the two systems,
equilibrium states are canonically stable up to the turning point of
temperature and microcanonically stable up to the turning point of energy. We
note that the dimensionless
temperature and  the dimensionless energy in the two models are related by
\begin{equation}
\eta=\sigma \beta m v_0^2\quad {\rm and}\quad 
\Lambda=-\frac{1}{\sigma} \frac{E}{Mv_0^2}\quad {\rm
where}\quad \sigma=\frac{GM}{Rv_0^2}.
\end{equation} 
In practice $\sigma\sim 1$ since $v_0$ corresponds to the virial velocity of
the cluster. We
see that the orders of magnitude of the dimensionless energies and
dimensionless temperatures
in Figs.
\ref{etalambda} and \ref{ebAintro} are consistent with each other.

\subsection{General relativity}

\begin{figure}
\begin{center}
\includegraphics[clip,scale=0.3]{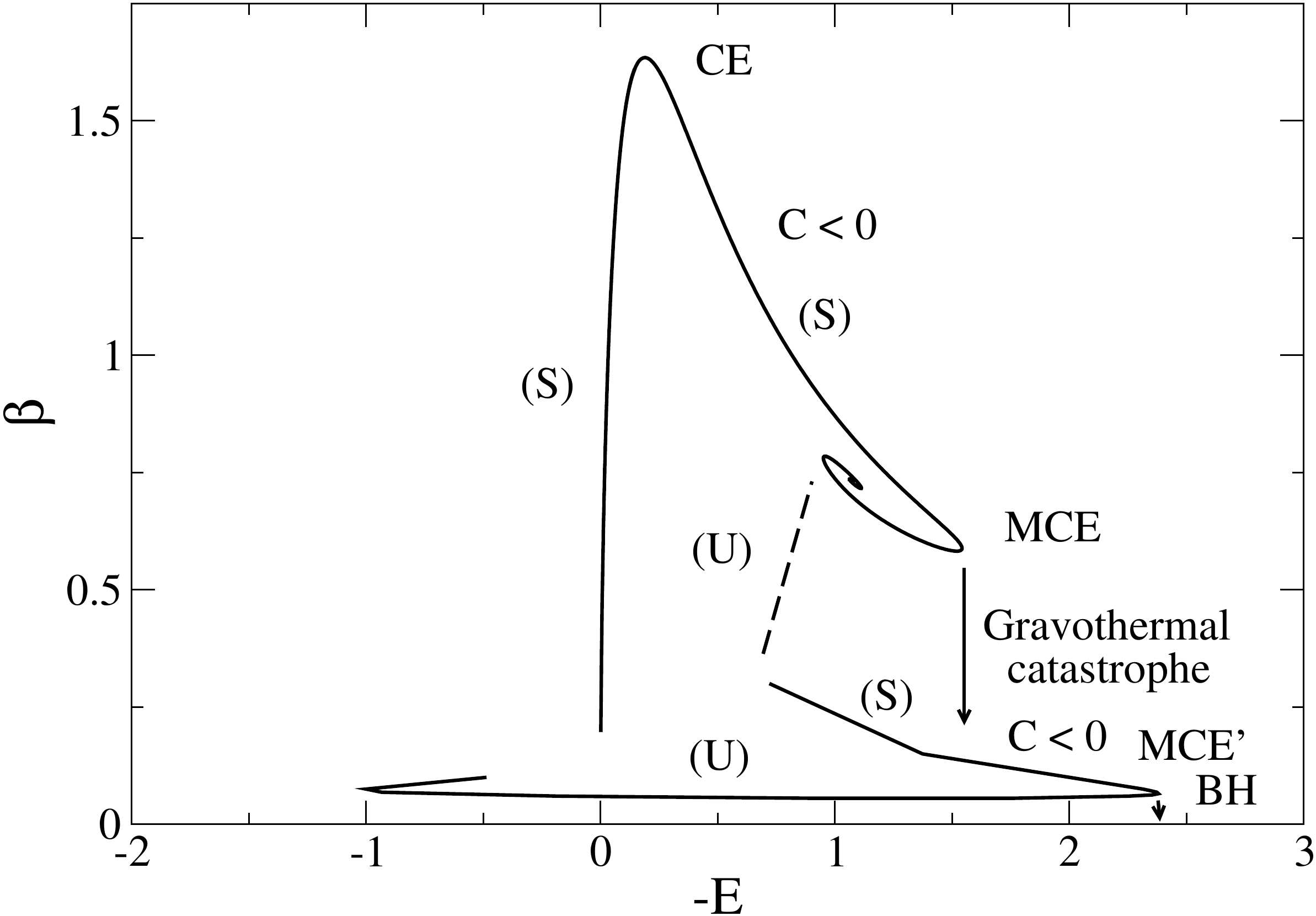}
\caption{Complete caloric curve of truncated isothermal star
clusters showing
the nonrelativistic cold spiral and the relativistic hot spiral (for
illustration we have taken $\nu=0.015$). The units of $\beta$ and $E$ are the
same as in Fig. \ref{ebAintro}. Nonrelativistic clusters undergo a gravothermal
catastrophe at MCE and
become relativistic. They can then undergo a catastrophic collapse at MCE'
leading to the formation of a black hole.}
\label{caloriccomplete}
\end{center}
\end{figure}

We now turn to the general relativistic case. The values of $E$ and $T_{\infty}$
tabulated by Ipser \cite{ipser69b} correspond
to strongly relativistic
stellar systems. Therefore, the caloric curve of Fig. \ref{ipser} describes only
the strongly relativistic part of the caloric curve (hot
spiral). We need to complete this caloric curve with the nonrelativistic part
from Fig. \ref{ebAintro} (cold
spiral). To that purpose, we first recall that, in the
nonrelativistic
regime, $\beta m v_0^2$ and ${E}/{Mv_0^2}$ are of the same order of magnitude as
$\eta$ and $\Lambda$ in the box model (see Appendix \ref{sec_sccnrng}). On the
other hand, in the
relativistic
regime, we have the relations
\begin{equation}
\eta=\nu\frac{mc^2}{k_B T_{\infty}},\quad {\rm and} \quad
\Lambda=-\frac{1}{\nu} \frac{E}{Nmc^2}\quad {\rm
where}\quad \nu=\frac{GNm}{Rc^2}.
\end{equation} 
As shown in Refs. \cite{roupas,acb}, the parameter $\nu$ belongs to the
interval $[0,0.1764]$.  In order to construct the complete
caloric curve of truncated isothermal star clusters, we
first determine the relativistic curve $\eta(\Lambda)$ from Fig. \ref{ipser} by
multiplying 
${mc^2}/{k_B T_{\infty}}$ by $\nu$ and dividing ${E}/{Nmc^2}$ by $\nu$, then we
add the nonrelativistic caloric curve of Fig. \ref{ebAintro}. We note that the
result depends on the parameter $\nu$ so we actually have a family
of caloric curves. For
illustration, we have taken $\nu=0.015$. This leads to the complete
caloric curve reported in Fig. \ref{caloriccomplete}.\footnote{We recall that
the isothermal model studied by
Zel'dovich and Podurets \cite{zp} and Isper  \cite{ipser69b} is  based on a
certain {\it ad hoc} relation between the 
energy cutoff and the 
temperature (see footnote 21). This is why their series of equilibria is unique.
If we relax this assumption, we get a family of
caloric curves as investigated by
\cite{khk,sf76,fs76,mr,mreuro,mr90,bmrv,bmrv2}. For box-confined
isothermal models \cite{roupas,acb}, we also have a family of caloric curves
parametrized by the compactness parameter $\nu=GNm/Rc^2$. These considerations 
explain why the
complete caloric curve of Fig. \ref{caloriccomplete} depends on a parameter
$\nu$.} Our procedure is of
course very approximate but it is sufficient to show the idea. It will be
important in future works to improve this procedure  and determine the family
of caloric curves exactly. 

From the complete caloric curve of Fig. \ref{caloriccomplete} it is now easy
to
understand the thermodynamical stability/instability of isothermal clusters in
both nonrelativistic and relativistic regimes. We again advocate the Poincar\'e
turning point criterion \cite{poincare}.

In the microcanonical ensemble, the nonrelativistic
branch is stable up to the first fundamental turning point of energy MCE (see
Appendix \ref{sec_sccnr}). At that point the series of equilibria
$\beta(-E)$  rotates clockwise so that a mode of stability is lost. A
new mode of stability is lost at each subsequent turning
point of energy where the series of equilibria rotates clockwise. At some point
the series of equilibria unwinds and rotates anti-clockwise (the unwinding of
the spiral is represented schematically by a dashed line in Fig.
\ref{caloriccomplete}). A mode of stability is gained at each turning point of
energy where the series of equilibria rotates anticlockwise. Finally, the
series
of equilibria on the relativistic branch becomes stable again (after rotating
anticlockwise an even number of times) and remains stable
until the second fundamental turning point of energy MCE' at which it becomes
unstable again. After that point the
series of equilibria rotates only clockwise so that it remains
unstable until the end. This is
consistent with the results from Appendix \ref{sec_sccr}.

In the canonical ensemble, the nonrelativistic
branch is stable up to the first turning point of temperature CE (see
Appendix \ref{sec_sccnr}). At that
point the series of equilibria $\beta(-E)$ rotates clockwise so that a mode
of stability is lost. The series of equilibria thus becomes unstable. It remains
unstable until the end because it can never recover the first mode of stability
that it has lost (it
rotates anticlockwise an odd number of times). This is why we found in Appendix
\ref{sec_sccr}
that the
relativistic branch is always unstable in the canonical ensemble.

The caloric curve of  Fig. \ref{caloriccomplete} provides a nice illustration
of the scenario discussed in the Introduction. Because of collisions and
evaporation a Newtonian stellar system evolves along the upper branch of the
series of equilibria. When it reaches the first fundamental turning point of
energy it becomes thermodynamically unstable (while remaining dynamically
stable) and undergoes a gravothermal catastrophe. It then evolves towards
very hot and very dense configurations and becomes relativistic. It
may then reach a stable relativistic isothermal equilibrium distribution. Again,
because of
collisions and evaporation, a relativistic star cluster evolves along the
lower branch of the series of equilibria. When it reaches the second fundamental
turning point of energy it becomes thermodynamically and dynamically unstable
and undergoes a catastrophic collapse towards a black hole.\footnote{It
is interesting to note that the
caloric curve of Fig. \ref{caloriccomplete} for classical truncated isothermal
star clusters
resembles the caloric curve in Fig. 29 of
\cite{acf} for self-gravitating fermions confined within a box in
general
relativity.}

We can now compare the caloric curve of truncated isothermal star
clusters (see Fig. \ref{caloriccomplete}) with the caloric curve of box-confined
isothermal systems (see Fig. \ref{kcal_N01_linked_colorsPH}). As we have seen
in Appendix \ref{sec_sccnrng}, the nonrelativistic parts of the caloric curves
are relatively similar. In particular, they present a ``cold'' spiral at
negative energies. We now consider the relativistic parts of the caloric curves.
In that case, a ``hot'' spiral arises because, in the ultrarelativistic limit 
(corresponding to very high energy densities), the equation of state of an
isothermal gas is $P=\epsilon/3$ like for the self-gravitating black-body
radiation.  However,
in the relativistic domain, the series of equilibria of truncated isothermal
distributions (see Figs. \ref{ipser} and \ref{caloriccomplete}) and the series
of equilibria of
box-confined isothermal systems (see Figs. \ref{kcal_N01_linked_colorsPH} and
\ref{hot}) are
very different. 

(i) The origin of the difference stems from the
fact that, for open clusters, stable equilibrium states necessarily have a
negative energy $E<0$.\footnote{In general
relativity, equilibrium states of star clusters may have a positive energy
while
this is not possible in Newtonian gravity (the equilibrium virial
theorem $2K+W=0$ implies $E=K+W=(1/2)W=-K<0$). This observation was first made
by Zel'dovich \cite{zeldovichEpos}. However, the
equilibrium states with
$E>0$ are generally unstable.}
By contrast, when the system is confined within a box, stable equilibrium states
with a positive energy are allowed because of the pressure exerted by the box.
For box-confined systems,  the ``hot
spiral'' appears at very large positive energies. At such energies, the
relativistic isothermal gas behaves as a form
of radiation confined within a cavity like in the studies of
Sorkin {\it et al.} \cite{sorkin} and Chavanis \cite{aarelat2}. 
There are no such equilibrium states with very large positive energies
in open star clusters. In open star clusters, the ``hot
spiral'' appears at energies (positive and negative) around $E=0$. In addition,
the stable equilibrium states necessarily have a negative energy.

(ii) For box-confined isothermal systems the cold and hot spiral are quite
distinct. The cold spiral appears at negative energies and the hot spiral appear
at large positive energies. For open star clusters, the cold and hot spirals
both appear at approximately the same range of (negative) energies. 
When general relativity is taken into account, the cold spiral of
Fig. \ref{ebAintro} unwinds
and connects the hot spiral of Fig. \ref{ipser} as sketched in Fig.
\ref{caloriccomplete}.

As a
result, a relativistic isothermal cluster in a box
behaves very differently from a truncated relativistic isothermal cluster. This
remark may question the physical relevance of the ``hot
spiral'' at large energies for realistic isothermal star clusters except if one
can justify a form of confinement playing the role of the box. This may
be the case for relativistic stars. The box
could be caused by a medium exerting a tremendously large pressure on an
ultrarelativistic gas, like in the supernova phenomenon where the system
achieves a ``core-halo'' structure. The core could correspond to the
ultrarelativistic isothermal gas and the halo could play the role of the box.
In that case, the core would be sustained by the external pressure of
the halo. However, this scenario remains to be put on a more rigorous basis.

\section{Entropy and free energy as functionals of the density for
nonrelativistic self-gravitating classical particles}
\label{sec_zalt}

We consider a nonrelativistic system of self-gravitating
classical particles.
The statistical
equilibrium state is
obtained by maximizing the Boltzmann entropy at fixed energy and particle number
in the
microcanonical ensemble, or
by minimizing the free energy at fixed particle number in the canonical
ensemble.  In Appendix C.2 of Paper I,  we have introduced entropy and free
energy functionals of the
distribution
function $f({\bf
r},{\bf v})$. In Sec. \ref{sec_gmnr} of this paper and in Appendix C.1.a of
Paper I, we have
introduced entropy and free energy functionals of the local density $n(r)$ and
local kinetic energy
$\epsilon_{\rm
kin}(r)$. In this Appendix, we introduce entropy and free energy
functionals
of the local density $n(r)$.

\subsection{Microcanonical ensemble}
\label{sec_altmce}

In the microcanonical ensemble,  the statistical equilibrium state is
obtained by maximizing the Boltzmann entropy $S[f]$ at fixed energy $E$ and 
particle number $N$. To solve this
maximization problem, we proceed in two steps. We first maximize 
$S[f]$ at fixed $E$, $N$
{\it and} particle density $n(r)$. Since $n(r)$ determines the particle number
$N[n]$ and the gravitational energy $W[n]$,
this is equivalent to maximizing $S[f]$ at fixed kinetic energy $E_{\rm
kin}[f]$ and particle density 
$n(r)$. The variational problem
for the first variations (extremization) can be written as
\begin{equation}
\label{alt2}
\frac{\delta S}{k_B}-\beta\delta E_{\rm kin}+\int\alpha(r)\delta n \, d{\bf
r}=0,
\end{equation}
where $\beta$ is a global (uniform) Lagrange multiplier and $\alpha(r)$ is
a local (position dependent) Lagrange multiplier. This variational problem 
yields 
\begin{equation}
\label{a2ann}
f({\bf r},{\bf p})=\frac{g}{h^3}e^{-\beta p^2/2m+\alpha(r)}.
\end{equation}
As in Appendix C1 of Paper I, we can show that this
distribution is the global maximum of
$S[f]$  at fixed $E_{\rm kin}[f]$  and $n(r)$. Substituting Eq.
(\ref{a2ann}) into Eqs. (I-17),
(I-18) and (I-20), we get
\begin{equation}
\label{za4}
n(r)=\frac{g}{h^3}( 2\pi m k_B
T)^{3/2}e^{\alpha(r)},
\end{equation}
\begin{equation}
\label{za5}
\epsilon_{\rm kin}(r)=\frac{3}{2}n(r)k_B T,
\end{equation}
\begin{equation}
\label{za6}
P(r)=\frac{2}{3}\epsilon_{\rm kin}(r)=n(r)k_B T.
\end{equation}
The Lagrange
multiplier $\alpha(r)$ is determined by the density $n(r)$ according to Eq.
(\ref{za4}). As a result, the distribution
function (\ref{a2ann}) can
be written in terms of the density as
\begin{equation}
\label{za17}
f({\bf r},{\bf p})=n({r}) \left (\frac{\beta}{2\pi m}\right
)^{3/2} e^{-\beta {p}^2/2m}.
\end{equation}
On the other hand, since $T$ is uniform, we see from Eq. (\ref{za6}) that the
equation of state is
isothermal: $P(r)=n(r) k_B T$. The temperature $T$ is determined by the
kinetic energy $E_{\rm kin}[n(r),T]=E-W[n(r)]$ using Eq. (\ref{za5})
integrated over the volume giving
$E_{\rm
kin}=(3/2)Nk_B T$. In other words, the temperature is determined by the energy
constraint 
\begin{eqnarray}
\label{alt22}
E=\frac{3}{2}Nk_B T+W[n(r)].
\end{eqnarray}
We note that $T$ is a functional of the density $n(r)$ but, for brevity, we
shall not write this dependence explicitly.  Substituting the
Maxwell-Boltzmann distribution (\ref{za17})
into
the entropy density (\ref{a1}) we obtain
\begin{equation}
\label{a8}
\frac{s(r)}{k_B}=\frac{5}{2}n(r)-n(r)\ln n(r)-n(r)\ln \left (\frac{h^3}{g}\right
)+\frac{3}{2}n(r)\ln (2\pi m k_B
T),
\end{equation}
which is equivalent to the integrated Gibbs-Duhem relation (\ref{a7}).
The entropy is given by
\begin{equation}
\label{za19}
\frac{S}{k_B}=\frac{5}{2}N-\int n(r)\ln n(r)\, 4\pi r^2\, dr-N\ln \left
(\frac{h^3}{g}\right
)+\frac{3}{2}N\ln (2\pi m k_B
T).
\end{equation}
Finally, the statistical equilibrium state in the microcanonical ensemble is
obtained by maximizing the entropy $S[n(r),T]$ at fixed particle
number $N$, the
energy constraint being taken into account in the determination of the
temperature  $T[n]$
through Eq. (\ref{alt22}). The variational problem
for the first variations (extremization) can be written as
\begin{equation}
\label{alt18b}
\frac{\delta S}{k_B}+\alpha_0\delta N=0.
\end{equation}
The conservation of energy implies [see Eq. (\ref{alt22})]:
\begin{eqnarray}
\label{alt22b}
0=\delta E_{\rm kin}+\int m\Phi\delta n\, d{\bf r}.
\end{eqnarray}
Using Eqs. (\ref{alt2}) and  (\ref{alt22b}), and proceeding as in Appendix F1
of Paper I, or performing the variations over $\delta n$ and $\delta T$
directly on the explicit expressions (\ref{alt22}) and (\ref{za19}) we obtain
\begin{equation}
\label{alt18d}
\alpha(r)=\alpha_0-\beta m\Phi(r)\qquad {\rm or}\qquad n(r)=\frac{g}{h^3}( 2\pi
m
k_B
T)^{3/2}e^{\alpha_0-\beta m\Phi(r)}.
\end{equation}
We then recover all the results of Sec. \ref{sec_smnrf}. The interest of this
formulation it that it allows us to solve more easily the stability problem
related
to
the sign of the second variations of entropy. This problem  has
been studied in detail in \cite{paddyapj,paddy,sc}. It has also been studied in
\cite{aarelat1} within the framework of special
relativity.

\subsection{Canonical ensemble}
\label{sec_altce}

In the canonical ensemble,  the statistical equilibrium state is
obtained by minimizing the Boltzmann free energy $F[f]=E[f]-TS[f]$ at fixed
particle
number $N$, or equivalently, by maximizing the Massieu function  
$J[f]=S[f]/k_B-\beta E[f]$ at fixed  particle number $N$. To solve this
maximization problem, we proceed in two steps. We first maximize 
$J[f]=S[f]/k_B-\beta E[f]$ at fixed $N$
{\it and} particle density $n(r)$. Since $n(r)$ determines the particle
number $N[n]$ and the gravitational energy $W[n]$,
this is equivalent to maximizing $S[f]/k_B-\beta E_{\rm kin}[f]$  at fixed
particle density $n(r)$. The variational problem
for the first variations (extremization) can be written as
\begin{equation}
\label{calt1}
\delta \left (\frac{S}{k_B}-\beta E_{\rm
kin}\right )+\int\alpha(r)\delta n \, d{\bf
r}=0,
\end{equation}
where $\alpha(r)$ is a local (position dependent)  Lagrange multiplier.
Since $\beta$ is constant in the canonical ensemble, this is equivalent to the
condition (\ref{alt2}) yielding the distribution function
(\ref{a2ann}). This distribution
is the global  maximum of
$S[f]/k_B-\beta E_{\rm kin}[f]$  at fixed $n(r)$. We then obtain the
same results as in Appendix \ref{sec_altmce}, except that  $T$ is fixed
while it was previously determined by the conservation of energy.

We can now simplify the expression of the free energy. The
entropy is given by Eq. (\ref{za19}) and the energy by Eq. (\ref{alt22}). 
Since $F=E-TS$, we obtain 
\begin{equation}
\label{ca22}
F=W[n(r)]-Nk_B T+k_B T\int n(r)\ln n(r)\, 4\pi r^2\, dr+Nk_BT\ln \left
(\frac{h^3}{g}\right
)-\frac{3}{2}Nk_BT\ln (2\pi m k_B
T).
\end{equation}
Using the foregoing results, the free energy can be written as a functional of
the density as
\begin{equation}
\label{calt13}
F[n(r)]=U[n(r)]+W[n(r)],
\end{equation}
where $U[n(r)]$ is the internal energy given by 
\begin{equation}
\label{calt14}
U[n(r)]=-Nk_B T+k_B T\int n(r)\ln n(r)\, 4\pi r^2\, dr+Nk_BT\ln \left
(\frac{h^3}{g}\right
)-\frac{3}{2}Nk_BT\ln (2\pi m k_B
T).
\end{equation}
Finally, the statistical equilibrium state in the canonical ensemble is
obtained by minimizing the free energy $F[n]$ at fixed particle
number $N$. The variational problem
for the first variations (extremization) can be written as
\begin{equation}
\label{alt19b}
\delta J+\alpha_0\delta N=0.
\end{equation}
Decomposing the Massieu function as $J[f]=S[f]/k_B-\beta E_{\rm
kin}[f]-\beta W[n]$ and proceeding as in Appendix F2
of Paper I, or performing the variations over $\delta n$ 
directly on the explicit expression (\ref{ca22}) of the free energy we get
\begin{equation}
\label{calt18d}
\alpha(r)=\alpha_0-\beta m\Phi(r)\qquad {\rm or}\qquad n(r)=\frac{g}{h^3}( 2\pi
m
k_B
T)^{3/2}e^{\alpha_0-\beta m\Phi(r)}.
\end{equation}
We then recover all the results of Sec. \ref{sec_smnrf}. The interest of this
formulation it that it allows us to solve more easily the stability problem
related
to
the sign of the second variations of free energy. This
problem has
been studied in detail in \cite{aaiso,sc}. It has also been studied in
\cite{aarelat1} within the framework of special
relativity.

{\it Remark:} We note that the free energy $F[\rho]$ coincides with the energy
functional associated with the Euler-Poisson equations describing a gas with an
isothermal equation of state $P=\rho k_B T/m$ (see Appendix G1 of Paper I). As a
result, the thermodynamical
stability of a classical self-gravitating system in the canonical ensemble is
equivalent
to the dynamical stability of the corresponding isothermal gas described by
the Euler-Poisson equations \cite{aaiso}.  This is a particular case of the
general result established in
\cite{aaantonov} which is valid for an arbitrary form of entropy. According to
the Poincar\'e turning point criterion, the series of
equilibria
becomes both thermodynamically unstable  (in the canonical ensemble) and
dynamically
unstable with respect to the Euler-Poisson equations at the first turning point
of temperature (or, equivalently, at the first turning point of mass).


\begin{thebibliography}{99}




\bibitem{theory1}{\small P.H. Chavanis, preprint (Paper I)}


\bibitem{tolman}{\small  R.C. Tolman, Phys. Rev. {\bf
35}, 904 (1930)}
\bibitem{cocke}{\small  W.J. Cocke, Ann. I.H.P.  {\bf
4}, 283 (1965)}
\bibitem{hk0}  {\small G. Horwitz, J. Katz, Ann. Phys. (USA) {\bf 76}, 301
(1973)} 
\bibitem{kh}  {\small J. Katz, G. Horwitz,  Astrophys.
J. {\bf 194}, 439 (1974)} 
\bibitem{khk}  {\small J. Katz, G. Horwitz, M. Klapisch, Astrophys. J. 
{\bf 199}, 307 (1975)}
\bibitem{kam}{\small  J. Katz, Y. Manor, Phys. Rev. D {\bf
12}, 956 (1975)}
\bibitem{kh2}  {\small J. Katz, G. Horwitz, Astrophys. J. 
{\bf 33}, 251 (1977)} 
\bibitem{hk}  {\small G. Horwitz, J. Katz, Astrophys.
J. {\bf 223}, 311 (1978)} 
\bibitem{ipser80}{\small J.R. Ipser, Astrophys. J. {\bf 238}, 1101 (1980)}
\bibitem{sorkin}{\small R.D. Sorkin, R.M. Wald, Z.Z. Jiu, Gen.
Relat. Grav. {\bf 13}, 1127 (1981)}
\bibitem{sy}{\small W.M. Suen, K. Young, Phys. Rev. A  {\bf
35}, 406 (1987)}
\bibitem{sy2}{\small W.M. Suen, K. Young, Phys. Rev. A  {\bf
35}, 411 (1987)}
\bibitem{bvrelat}{\small N. Bilic, R.D. Viollier, Gen. Relat. Grav. {\bf
31}, 1105 (1999)}
\bibitem{aarelat1}  {\small P.H. Chavanis, Astron. Astrophys.  {\bf 381}, 709
(2002)}
\bibitem{aarelat2}{\small P.H. Chavanis, Astron. Astrophys. {\bf
483}, 673 (2008)}
\bibitem{gao}{\small S. Gao, Phys. Rev. D {\bf
84}, 104023 (2011)}
\bibitem{gaoE}{\small S. Gao, Phys. Rev. D {\bf
85}, 027503 (2012)}
\bibitem{roupas1}{\small Z. Roupas, Class. Quantum Grav. {\bf
30}, 115018 (2013)}
\bibitem{gsw}  {\small S.R. Green, J.S. Schiffrin, R.M. Wald, Class. Quantum
Grav. {\bf 31}, 035023 (2014)}
\bibitem{fg}  {\small X. Fang, S. Gao, Phys. Rev. D {\bf
90}, 044013 (2014)}
\bibitem{roupas1E}{\small Z. Roupas, Class. Quantum Grav. {\bf
32}, 119501 (2015)}
\bibitem{schiffrin}{\small J.S. Schiffrin, Class. Quantum Grav.  {\bf 32},
185011 (2015)}
\bibitem{psw}{\small K. Prabhu, J.S. Schiffrin, R.M. Wald, Class. Quantum
Grav. {\bf 33},
185007 (2016)}
\bibitem{fhj}{\small X. Fang, X. He, J. Jing, Eur. Phys. J. C  {\bf
77}, 893 (2017)}
\bibitem{ov}{\small J.R. Oppenheimer, G.M. Volkoff, Phys. Rev. {\bf 55}, 374
(1939)}
\bibitem{klein}{\small  O. Klein, Rev. Mod. Phys. {\bf
21}, 531 (1949)}
\bibitem{acf}{\small G. Alberti, P.H. Chavanis, arXiv:1808.01007}
\bibitem{acb}{\small G. Alberti, P.H. Chavanis, arXiv:1908.10316}
\bibitem{clm1}{\small P.H. Chavanis, M. Lemou, F. M\'ehats, Phys. Rev. D {\bf
91}, 063531 (2015)}
\bibitem{chandrabook}  {\small S. Chandrasekhar, {\it Principles of Stellar
Dynamics} (University of Chicago Press, 1942)}
\bibitem{ambartsumian}  {\small V.A. Ambartsumian, Ann. Leningrad
State Univ. {\bf 22}, 19 (1938)}
\bibitem{spitzer}  {\small L. Spitzer, MNRAS  {\bf 100}, 396 (1940)}
\bibitem{woolley}  {\small  R. Woolley, MNRAS {\bf 114}, 191 (1954)}
\bibitem{king}{\small I. King, Astron. J. {\bf 71}, 64 (1966)}
\bibitem{eddington3}  {\small  A.S. Eddington, MNRAS {\bf 76}, 572 (1916)}
\bibitem{michie}  {\small  R.W. Michie, MNRAS {\bf 125}, 127 (1963)}
\bibitem{lbw}  {\small  D. Lynden-Bell, R. Wood, Mon. Not. R. Astron. Soc.
{\bf 138}, 495 (1968)}
\bibitem{katzking}  {\small  J. Katz, MNRAS {\bf 190}, 497 (1980)}
\bibitem{antonov}  {\small V.A. Antonov,  Vest. Leningr. Gos. Univ. {\bf 7}, 135
(1962)}
\bibitem{larson1}  {\small R.B. Larson, Mon. Not. R. Astron.
Soc. {\bf 147}, 323 (1970)}
\bibitem{larson2}  {\small R.B. Larson, Mon. Not. R. Astron.
Soc. {\bf 150}, 93 (1970)}
\bibitem{hmc}  {\small M. H\'enon, Astrophys. Space Sci. {\bf 13}, 284 (1971)}
\bibitem{hnns}  {\small I. Hachisu, Y. Nakada, K. Nomoto, D. Sugimoto, Prog.
Theor. Phys. {\bf 60}, 393 (1978)}
\bibitem{lbe}  {\small D. Lynden-Bell, P.P. Eggleton, Mon. Not. R. Astron. Soc.
{\bf 191}, 483 (1980)}
\bibitem{cohn} {\small H. Cohn, Astrophys. J. {\bf 242}, 765 (1980)}
\bibitem{inagakilb}  {\small S. Inagaki, D. Lynden-Bell, Mon. Not. R. Astron.
Soc.
{\bf 205}, 913 (1983)}
\bibitem{oscillations} {\small D. Sugimoto, E. Bettwieser,   Mon. Not. R.
Astron. Soc. {\bf 204}, 19 (1983)}
\bibitem{hr} {\small D. Heggie, N. Ramamani, Mon. Not. R. Astron. Soc. {\bf
237}, 757 (1989)}
\bibitem{doremus71}  {\small J.P. Doremus, M.R. Feix, G. Baumann,
Phys. Rev. Lett. {\bf 26}, 725 (1971)}
\bibitem{doremus73}  {\small J.P. Doremus, M.R. Feix, G. Baumann,
Astron. Astrophys. {\bf 29}, 401 (1973)}
\bibitem{gillon76}  {\small D. Gillon, M. Cantus, J.P. Doremus, G.
Baumann, Astron. Astrophys. {\bf 50}, 467 (1976)}
\bibitem{sflp}  {\small J.F. Sygnet, G. Des Forets, M. Lachieze-Rey, R. Pellat,
Astrophys. J. {\bf 276}, 737 (1984)}
\bibitem{ks}  {\small H. Kandrup, J.F. Sygnet, Astrophys. J. {\bf 298}, 27
(1985)}
\bibitem{kandrup91}{\small H. Kandrup, Astrophys. J. {\bf 370}, 312
(1991)}
\bibitem{zp}  {\small Y.B. Zel'dovich, M.A. Podurets, Soviet Astron. -- AJ  {\bf
9}, 742 (1966)}
\bibitem{fackerellthesis}  {\small E.D. Fackerell, Ph.D. thesis, University of
Sydney (1966)}
\bibitem{fackerell}{\small E. Fackerell, Astrophys. J. {\bf 153},
643 (1968)}
\bibitem{ipserthorne}{\small J.R. Ipser, K.S. Thorne, Astrophys. J. {\bf 154},
251 (1968)}
\bibitem{ipser69}{\small J.R. Ipser, Astrophys. J. {\bf 156}, 509
(1969)}
\bibitem{ipser69b}{\small J.R. Ipser, Astrophys. J. {\bf 158}, 17
(1969)}
\bibitem{fackerell70}{\small E. Fackerell, Astrophys. J. {\bf 160},
859 (1970)}
\bibitem{sudbury} {\small A.W. Sudbury, Mon. Not. R. Astron. Soc. {\bf
147}, 187 (1970)}
\bibitem{sf76}  {\small K. Suffern, E. Fackerell, Astrophys. J. {\bf 203}, 477
(1976)}
\bibitem{fs76}  {\small E. Fackerell, K. Suffern,  Aust. J. Phys. {\bf 29}, 311
(1976)}
\bibitem{mr}  {\small M. Merafina, R. Ruffini, Astron.
 Astrophys. {\bf 221}, 4 (1989)}
\bibitem{mreuro}  {\small M. Merafina, R. Ruffini, Europhys. Lett. {\bf 9}, 621
(1989)}
\bibitem{mr90}  {\small M. Merafina, R. Ruffini, Astron.
 Astrophys. {\bf 227}, 415 (1990)}
\bibitem{bmrv}  {\small G.S. Bisnovatyi-Kogan, M. Merafina, R. Ruffini, E.
Vesperini, Astrophys. J. {\bf 414}, 187 (1993)}
\bibitem{bmrv2}  {\small G.S. Bisnovatyi-Kogan, M. Merafina, R. Ruffini, E.
Vesperini, Astrophys. J. {\bf 500}, 217 (1998)}
\bibitem{poincare}  {\small H. Poincar\'e, Acta Math. {\bf 7}, 259 (1885)}
\bibitem{fit}  {\small D. Fackerell, J. Ipser, K. Thorne, Comments Ap. and Space
Phys. {\bf 1}, 134 (1969)}
\bibitem{st1} {\small S.L. Shapiro, S.A. Teukolsky,  Astrophys. J.
{\bf 298}, 34 (1985)}
\bibitem{st2} {\small S.L. Shapiro, S.A. Teukolsky,  Astrophys. J.
{\bf 298}, 58 (1985)}
\bibitem{st3} {\small S.L. Shapiro, S.A. Teukolsky,  Astrophys. J. 
{\bf 292}, L41 (1985)}
\bibitem{st4} {\small S.L. Shapiro, S.A. Teukolsky,  Astrophys. J. 
{\bf 307}, 575 (1986)}
\bibitem{kochanek} {\small C.S. Kochanek, S.L. Shapiro, S.A. Teukolsky, 
Astrophys. J. {\bf 320}, 73 (1987)}
\bibitem{rst1} {\small F. Rasio, S.L. Shapiro, S.A. Teukolsky, 
Astrophys. J. {\bf 344}, 146 (1989)}
\bibitem{strevue} {\small S.L. Shapiro, S.A. Teukolsky, Phil. Trans. R. Soc.
Lond. A {\bf 340}, 365 (1992)}
\bibitem{stgas1} {\small S.L. Shapiro, S.A. Teukolsky, 
Astrophys. J. {\bf 234}, L177 (1979)}
\bibitem{stgas2} {\small S.L. Shapiro, S.A. Teukolsky, 
Astrophys. J. {\bf 235}, 199 (1980)}
\bibitem{hf67}  {\small F. Hoyle, W.A. Fowler, Nature {\bf
213}, 373 (1967)}
\bibitem{hf63a}  {\small F. Hoyle, W.A. Fowler, Mon. Not. R. Astron. Soc.  {\bf
125}, 169 (1963)}
\bibitem{hf63b}  {\small F. Hoyle, W.A. Fowler, Nature {\bf
197}, 533 (1963)}
\bibitem{zapolsky} {\small H.S. Zapolsky, Astrophys. J.
{\bf 153}, L163 (1968)}
\bibitem{salpeter} {\small E.E. Salpeter, Astrophys. J.
{\bf 140}, 796 (1964)}
\bibitem{zeldovichBH} {\small Ya.B. Zel'dovich, Soviet Phys. Doklady {\bf 9},
195 (1964)}
\bibitem{bkz}  {\small G.S. Bisnovatyi-Kogan, Ya. B. Zel'dovich,
Astrofizika {\bf 5}, 223 (1969)}
\bibitem{bkt}  {\small G.S. Bisnovatyi-Kogan, K.S. Thorne, Astrophys. J.
 {\bf 160}, 875 (1970)}
\bibitem{rst2} {\small F. Rasio, S.L. Shapiro, S.A. Teukolsky, 
Astrophys. J. {\bf 336}, L63 (1989)}
\bibitem{mrz}  {\small M. Merafina, R. Ruffini, Astrophys. J. {\bf 454}, L89
(1995)}
\bibitem{balberg} {\small S. Balberg, S.L. Shapiro, S. Inagaki, Astrophys. J.
{\bf 568}, 475 (2002)}
\bibitem{bash} {\small S. Balberg, S.L. Shapiro, Phys. Rev. Lett.
{\bf 88}, 101301 (2002)}
\bibitem{pss} {\small J. Pollack, D. Spergel, P. Steinhardt, Astrophys. J.
{\bf 804}, 131 (2015)}
\bibitem{clm2}{\small P.H. Chavanis, M. Lemou, F. M\'ehats, Phys. Rev. D {\bf
92}, 123527 (2015)}
\bibitem{bosons}{\small P.H. Chavanis, arXiv:1810.08948}
\bibitem{ipser74}{\small J.R. Ipser, Astrophys. J. {\bf 193},
463 (1974)}
\bibitem{tvh}  {\small L. Taff, H. van Horn, Astrophys.
J. {\bf 197}, L23 (1975)} 
\bibitem{hkpart1}  {\small G. Horwitz, J. Katz, Astrophys.
J. {\bf 211}, 226 (1977)} 
\bibitem{nakada}  {\small Y. Nakada, Publ. Astron. Soc. Japan
{\bf 30}, 57 (1978)}
\bibitem{hs}  {\small I. Hachisu, D. Sugimoto, Prog. Theor.
Phys. {\bf 60}, 123 (1978)}
\bibitem{hk3}  {\small G. Horwitz, J. Katz,  Astrophys. J. {\bf 222}, 941
(1978)}
\bibitem{katzpoincare1}  {\small J. Katz,  Mon. Not. R. Astron. Soc. {\bf 183},
765 (1978)}
\bibitem{ih}  {\small J.R. Ipser, G. Horwitz,  Astrophys. J.
{\bf 232}, 863 (1979)}
\bibitem{inagaki}  {\small S. Inagaki, Publ. Astron. Soc. Japan
{\bf 32}, 213 (1980)}
\bibitem{lecarkatz}  {\small  M. Lecar, J. Katz, Astrophys. J. {\bf 243}, 983
(1981)}
\bibitem{ms}  {\small J. Messer, H. Spohn, J. Stat. Phys.
{\bf 29}, 561 (1982)}
\bibitem{lp}  {\small  J.F. Luciani, R.
Pellat, Astrophys. J. {\bf 317}, 241 (1987)}
\bibitem{kiessling}  {\small M. Kiessling, J. Stat. Phys.
{\bf 55}, 203 (1989)}
\bibitem{paddyapj}{\small T. Padmanabhan, Astrophys. J. Supp. {\bf 
71}, 651 (1989)}
\bibitem{paddy}  {\small T. Padmanabhan,  Phys. Rep.  {\bf 188}, 285 (1990)}
\bibitem{dvsc}  {\small H.J. de Vega, N. Sanchez, F. Combes, Phys. Rev. D  {\bf
54}, 6008 (1996)}
\bibitem{semelin0}  {\small B. Semelin,  H.J. de
Vega, N. Sanchez, F. Combes, Phys. Rev. D {\bf 59}, 125021 (1999)}
\bibitem{katzokamoto}  {\small  J. Katz, I. Okamoto, MNRAS {\bf 317}, 163
(2000)}
\bibitem{semelin}  {\small B. Semelin, N. Sanchez, H.J. de
Vega, Phys. Rev. D {\bf 63}, 084005 (2001)}
\bibitem{dvs1}  {\small H.J. de Vega, N. Sanchez, Nucl. Phys. B  {\bf
625}, 409
(2002)}
\bibitem{dvs2}  {\small H.J. de Vega, N. Sanchez, Nucl. Phys. B  {\bf 625}, 460
(2002)}
\bibitem{aaiso}  {\small P.H. Chavanis, Astron. Astrophys. {\bf 381}, 340
(2002)}
\bibitem{crs}{\small P.H. Chavanis, C. Rosier, C. Sire, Phys. Rev. E {\bf
66}, 036105 (2002)}
\bibitem{sc}{\small  C. Sire, P.H. Chavanis, Phys. Rev. E {\bf
66}, 046133 (2002)}
\bibitem{grand}  {\small P.H. Chavanis, Astron. Astrophys. {\bf 401}, 15
(2003)}
\bibitem{katzrevue}  {\small J. Katz, Found. Phys. {\bf 33}, 223 (2003)}
\bibitem{lifetime}  {\small P.H. Chavanis, Astron. Astrophys. {\bf 432}, 117
(2005)}
\bibitem{ijmpb}  {\small P.H. Chavanis, Int. J. Mod. Phys. B {\bf 20}, 3113
(2006) }
\bibitem{sb}  {\small M. Sormani, G. Bertin, Astron. Astrophys.
{\bf 552}, A37
(2013)}
\bibitem{roupas}{\small Z. Roupas, Class. Quantum Grav. {\bf 
32}, 135023 (2015)}
\bibitem{post}{\small C. Sire, P.H. Chavanis, Phys. Rev. E {\bf
69}, 066109 (2004)}
\bibitem{juttner1}{\small F. Juttner, Ann. Phys. {\bf 339}, 856 (1911)}
\bibitem{juttner2}{\small F. Juttner, Ann. Phys. {\bf 340}, 145 (1911)}
\bibitem{planck}{\small M. Planck, Ann. Phys. {\bf 26}, 1 (1908)}
\bibitem{aaantonov}  {\small P.H. Chavanis,  Astron. Astrophys. {\bf
451}, 109 (2006)}
\bibitem{cc}{\small A. Campa, P.H. Chavanis, J. Stat. Mech. {\bf 06},
06001 (2010)}
\bibitem{chandra72}  {\small S. Chandrasekhar, in
General Relativity, papers in honour of J.L. Synge, Edited by L.O' Raifeartaigh
(Oxford, 1972)}
\bibitem{pt}  {\small P.H. Chavanis, Phys. Rev. E {\bf 65}, 056123 (2002)}
\bibitem{dk}  {\small N.A. Dmitriev, S.A. Kholin, Voprosy kosmogonii {\bf
9}, 254 (1963)}
\bibitem{zeldovichEpos} {\small Ya.B. Zel'dovich, Soviet Phys. JETP {\bf 15},
1158 (1962)}








\end{thebibliography}
\end{document}